\documentclass[prc,aps,superscriptaddress,twocolumn,showpacs,nofootinbib]{revtex4}

\usepackage{graphicx,amsmath,amssymb,bm,multirow}
\usepackage{amsthm,mathrsfs}
\usepackage{amsfonts}
\usepackage{dcolumn}
\usepackage{bm}
\usepackage[colorlinks,linkcolor=blue,anchorcolor=blue,citecolor=blue]{hyperref}
\usepackage[usenames,dvipsnames]{color}

\newcommand{\br}{{\mathbf{r}}}

\newcommand{\balp}{{\boldsymbol{\alpha}}}  

\newcommand{\bnab}{{\mathbf{\nabla}}}

\newcommand{\beq}{\begin{equation}}
\newcommand{\eeq}{\end{equation}}
\newcommand{\beqn}{\begin{eqnarray}}
\newcommand{\eeqn}{\end{eqnarray}}

\newcommand{\bsub}{  \begin{subequations}}
\newcommand{\esub}{ \end{subequations}}

\renewcommand{\vec}[1]{\mbox{\boldmath $#1$}}

\begin{document}

\title{Microscopic study of low-lying spectra of  $\Lambda$ hypernuclei based on a beyond-mean-field approach with covariant energy density functional}

\author{H. Mei}
\affiliation{Department of Physics, Tohoku University, Sendai 980-8578,Japan}
\affiliation{School of Physical Science and Technology,
             Southwest University, Chongqing 400715, China}
\author{K. Hagino}
\affiliation{Department of Physics, Tohoku University, Sendai 980-8578,Japan}
\affiliation{Research Center for Electron Photon Science, Tohoku University, 1-2-1 Mikamine, Sendai 982-0826, Japan}
\affiliation{
National Astronomical Observatory of Japan, 2-21-1 Osawa,
Mitaka, Tokyo 181-8588, Japan}
\author{J. M. Yao}
\affiliation{Department of Physics, Tohoku University, Sendai 980-8578,Japan}
\affiliation{School of Physical Science and Technology,
             Southwest University, Chongqing 400715, China}
\author{T. Motoba}
\affiliation{Laboratory of Physics, Osaka Electro-Communications University,
             Neyagawa 572-8530, Japan}
\affiliation{Yukawa Institute for Theoretical Physics, Kyoto University, Kyoto 606-8502, Japan }

\begin{abstract}
We present a detailed formalism of the microscopic
particle-rotor model for hypernuclear low-lying states
based on a covariant density functional theory.
In this method, the hypernuclear states are constructed by
coupling a hyperon to low-lying states of the core nucleus,
which are described by the generator coordinate method (GCM)
with the particle
number and angular momentum projections.
We apply this method to study in detail the low-lying spectrum of
$^{13}_{~\Lambda}$C and $^{21}_{~\Lambda}$Ne hypernuclei.
We also briefly discuss the structure of
$^{155}_{~~\Lambda}$Sm as an example of heavy deformed hypernuclei.
It is shown that
the low-lying excitation spectrum with positive parity states
of the
hypernuclei, which are dominated by $\Lambda$ hyperon in $s$-orbital
coupled to the core states, are similar to
that for the corresponding core states, while
the electric quadrupole transition strength, $B(E2)$,
from the 2$^+_1$ state to the ground state is reduced
according to
the mass number of the hypernuclei.
Our study indicates that the energy splitting between
the first 1/2$^-$ and 3/2$^-$ hypernuclear states is
generally small for all the hypernuclei which we study.
However, their configurations depend much
on the properties of a core nucleus, in particular on the
sign of deformation parameter.
That is,
the first $1/2^-$ and $3/2^-$ states in
$^{13}_{~\Lambda}$C
are dominated by a single configuration with $\Lambda$ particle
in the $p$-wave orbits
and thus providing
good candidates for a study of the $\Lambda$ spin-orbit splitting.
On the other hand,
those states in the
other hypernuclei exhibit a large configuration mixing and thus their energy
difference cannot be interpreted as the spin-orbit splitting for
the $p$-orbits.
\end{abstract}
\maketitle

\section{Introduction}
The development in $\Lambda$-hypernuclear spectroscopy has enabled one to explore several
aspects of hypernuclear structure~\cite{Hashimoto06,Tamura09}.
Since hyperon-nucleon and hyperon-hyperon scattering experiments are difficult to
perform,
the information on the $\Lambda$N and $\Lambda\Lambda$ interactions have been extracted
from such studies.
Moreover, since a $\Lambda$ hyperon is free from the Pauli principle from other nucleons,
it can go deeply inside a nucleus, which can also be used
as a sensitive probe in order to study the structure
of normal nuclei.
Theoretically, many methods have been developed to investigate the
spectroscopy  of hypernuclei, such as the cluster model~\cite{Motoba83,Hiyama99,Bando90,Hiyama03,Cravo02, Suslov04, Shoeb09},
the shell model~\cite{Dalitz78,Gal71,Millener}, the ab-initio method~\cite{abinitio},
the antisymmetrized molecular dynamics (AMD)~\cite{Isaka11,Isaka11-2,Isaka12,Isaka13}, and
self-consistent mean-field models~\cite{Zhou07,Win08,Schulze10,Win11,Lu11,Weixia14,Li13,Lu14,HY14}.
Among them, the self-consistent mean-field approach is the only microscopic method which
can be globally applied from light to heavy hypernuclei.

One of the characteristic features of atomic nuclei is deformation in the
body-fixed frame. In the mean-field approach, the optimized deformation is
automatically obtained by minimizing the total energy of a system in the
mean-field approximation.
It was shown, however, that
the potential
energy surface of a hypernucleus is generally softer against deformation
than that of the corresponding core nucleus~\cite{Win11}.
This implies that the shape fluctuation effect, which is not
included in the pure mean-field approximation,
will be more important in hypernuclei than in normal nuclei.
Furthermore, in order to connect mean-field results to
spectroscopic observables, such as $B(E2)$ values,
one has to transform the results to the laboratory frame.
To this end,
one has to
rely on additional assumptions such as the rigid rotor model,
which however would not work for, {\it e.g.,}
nuclei with small deformation or with shape coexistence.
To quantify the impurity effect of $\Lambda$ particle
on nuclear structure,
one thus has to go beyond the pure
mean-field approximation.

In our previous publication~\cite{Mei14}, we have proposed
a new approach using a microscopic particle-rotor model (PRM)
for the low-lying states of single-$\Lambda$ hypernuclei.
In this method, the $\Lambda$ particle is coupled to the
core nucleus
states while the $\Lambda$ hyperon interacts with the nucleons
inside the nuclear core.
For the core nucleus,
a beyond-relativistic-mean-field approach is applied for low-lying states
by carrying out the angular momentum and the particle number projections
as well as the configuration mixing with the generator coordinate method
(GCM).
We have successfully applied this method to the 
spectrum of $^9_\Lambda$Be. 

The motivation of the present work is to introduce
a detailed formalism of this method. At the same time,
we also apply it systematically
in order to study the low-lying states of single-$\Lambda$
hypernuclei in the mass region from light to heavy.
For this purpose, we will first discuss
the $^{13}_{~\Lambda}$C hypernucleus, which is an ideal hypernucleus
in order to discuss the spin-orbit splitting.
We will then discuss $^{21}_{~\Lambda}$Ne
as an example of
hypernuclei in the $sd$-shell region with a prolate deformation.
The $^{155}_{~~\Lambda}$Sm hypernucleus, which is a well-deformed system
and has a well-developed ground-state rotational band,
will also be considered as an example of heavy hypernuclei.

The paper is organized as follows. In Sec.~\ref{Sec:framework} we present
a detailed introduction to the formalism of microscopic PRM
based on the covariant density functional theory (CDFT). 
In Sec. ~\ref{Sec:Results},
the results for the low-lying states of hypernuclei and
the corresponding core nuclei will be presented.
Finally,
we will summarize the conclusions of this paper
in Sec.~\ref{Sec:Summary}.

\section{Microscopic particle-rotor model for $\Lambda$ hypernuclei}
\label{Sec:framework}
\subsection{Coupled-channels equations}
\label{Sec:PRM}

\begin{figure}[]
\centering
 \includegraphics[width=7cm]{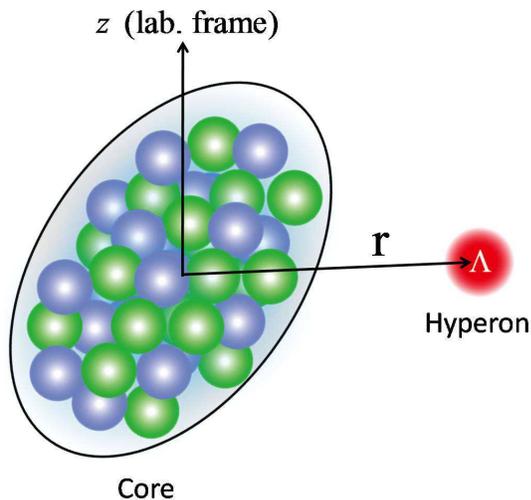}
 \caption{(Color online) A schematic picture of the microscopic particle-rotor model
for $\Lambda$ hypernucleus, in which $\vec{r}$ denotes the coordinate of the $\Lambda$
hyperon. In this approach, the nuclear core states are
described microscopically
with the multi-reference density functional theory.}
 \label{schematic}
 \end{figure}

The basic idea of the microscopic PRM for a single-$\Lambda$ hypernucleus is that the valence $\Lambda$ hyperon couples to the low-lying states of nuclear core,
as is illustrated in a schematic picture of Fig.~\ref{schematic}.
In this approach, a hypernucleus is described in the
laboratory frame and the wave function of the whole $\Lambda$ hypernucleus is constructed as
\begin{equation}
 \label{wavefunction}
 \displaystyle \Psi_{JM}(\vec{r},\{\vec{r}_N\})
 =\sum_{n,j,\ell, I}  {\mathscr R}_{j\ell I_{n}}(r) {\mathscr F}^{JM}_{j\ell I_{n}}(\hat{\vec{r}}, \{\vec{r}_N\}),
\end{equation}
where
\begin{equation}
 {\mathscr F}^{JM}_{j\ell I_{n}}(\hat{\vec{r}}, \{\vec{r}_N\})
= [{\mathscr Y}_{j\ell}(\hat{\vec{r}})\otimes
\Phi_{I_{n}}(\{\vec{r}_N\})]^{(JM)}
\end{equation}
with $\vec{r}$ and $\vec{r}_N$ being the coordinates of the $\Lambda$ hyperon and the
nucleons, respectively. $J$ is the angular momentum
for the whole system while $M$ is its
projection onto the $z$-axis.
${\mathscr Y}_{j\ell}(\hat{\vec{r}})$ is the spin-angular wave function for the $\Lambda$ hyperon.
$\vert\Phi_{I_{n}}\rangle$ is the wave functions of the low-lying states
of nuclear core, where $I$ represents the angular momentum
of the core state and $n=1, 2, \ldots$ distinguish
different core states with the same angular momentum $I$.  For convenience, hereafter we introduce the
shorthand notation $k=\{j\ell I_{n}\}$ to represent different channels.

In the relativistic approach, ${\mathscr R}_{k}(r)$ is the radial wave function of a four-component Dirac spinor and it can be written in the following form
\begin{equation}
\label{wavefunction2}
{\mathscr R}_{k}(r)=
\begin{pmatrix}
f_{k}(r)\\
i g_{k}(r)\vec{\sigma} \cdot \hat{\vec{r}}
\end{pmatrix}.
\end{equation}

The Hamiltonian $\hat H$ for the whole $\Lambda$ hypernucleus can be written as
\beq
\hat H = \hat T_\Lambda + \sum^{A_c}_{i=1} \left[\hat{V}_V^{(N\Lambda)}(\vec{r},\vec{r}_{N_i}) + \hat{V}_S^{(N\Lambda)}(\vec{r},\vec{r}_{N_i})\right] + \hat H_{\rm c},
\label{eq:H}
\eeq
where $A_c$ is the mass number of the core nucleus.
The first term in Eq. (\ref{eq:H}), $\hat T_\Lambda$,
is the kinetic energy of $\Lambda$ hyperon,
\beq
\hat T_\Lambda = -i \balp\cdot\bnab_\Lambda + \gamma^0 m_\Lambda.
\eeq
Here,
 $m_\Lambda$ is the mass of the $\Lambda$ hyperon and $\alpha$ and $\gamma^0$ are Dirac matrices. The second term in Eq. (\ref{eq:H})
represents the $N\Lambda$ interaction term between the
valence $\Lambda$ and the nucleons in the core nucleus.
It is composed of both a
repulsive vector-type term $\hat{V}_V^{(N\Lambda)}$ and an
attractive scalar-type term $\hat{V}_S^{(N\Lambda)}$, for which we take
the following contact coupling forms,
\begin{eqnarray}
\label{interaction}
\displaystyle \hat{V}_V^{(N\Lambda)}(\vec{r},\vec{r}_{N})&=& \alpha_V^{N\Lambda} \delta(\vec{r}-\vec{r}_{N}) \\
\label{interaction2}
\displaystyle \hat{V}_S^{(N\Lambda)}(\vec{r},\vec{r}_{N})&=& \alpha_S^{N\Lambda} \gamma^0_\Lambda \delta(\vec{r}-\vec{r}_{N})\gamma^0_N.
\end{eqnarray}
These $N\Lambda$ interactions correspond to the leading-order
four-fermion coupling terms in
the effective interaction proposed in Ref.~\cite{Tanimura2012}.
For simplicity, the possible higher-order derivative
and tensor coupling terms are neglected in the present study.
We note here that the spin-orbit interaction of hyperon is
automatically taken into account
in the relativistic framework
without introducing an additional parameter.
The last term in 
Eq. (\ref{eq:H}) 
is the many-body Hamiltonian for the core nucleus
satisfying
the equation  $\hat H_c \vert\Phi_{I_{n}}\rangle= E_{I_{n}} \vert\Phi_{I_{n}}\rangle$.
In this paper, we use $H_c$ that corresponds to
the following point-coupling energy density functional (EDF)~\cite{Buvenich02},
 \begin{widetext}
 \begin{eqnarray}\label{EDF}
 E_{\rm c}[\{\rho_i\}, \{j^\mu_i\}]
 &=&{\rm Tr}[(\mathbf{\alpha}\cdot\mathbf{p}+\beta m)\rho_V]
 + \int d{\bm r } \left(\frac{\alpha_S}{2}\rho_S^2+\frac{\beta_S}{3}\rho_S^3 +
    \frac{\gamma_S}{4}\rho_S^4+\frac{\delta_S}{2}\rho_S\triangle \rho_S
    +  \frac{\alpha_V}{2}j_\mu j^\mu + \frac{\gamma_V}{4}(j_\mu j^\mu)^2
    \right.\nonumber \\
   &&+\left.
     \frac{\delta_V}{2}j_\mu\triangle j^\mu
     +  \frac{\alpha_{TV}}{2}j^{\mu}_{TV}(j_{TV})_\mu+\frac{\delta_{TV}}{2}
    j^\mu_{TV}\triangle  (j_{TV})_{\mu}
     +\frac{e}{2}j^\mu_p A_\mu
 \right),
 \end{eqnarray}\end{widetext}
where
$A^\mu$ is the four-component electromagnetic field, and
the densities $\rho_i$ and currents  $j^\mu_i$ are bi-linear 
combinations of Dirac spinors, namely $\bar\psi\Gamma_i\psi$ 
with $i=S, V, TV$ representing the symmetry of the coupling. 
The subscript $S$ stands for isoscalar-scalar ($\Gamma_S = 1$), $V$ 
for isoscalar-vector  ($\Gamma_V = \gamma^\mu$), and $TV$ for 
isovector-vector ($\Gamma_{TV} =\gamma^\mu t_3$) type of coupling 
characterized by their transformation properties in isospin and 
in space-time.

In order to obtain
the radial wave function given by Eq. (\ref{wavefunction2})  and the
energy of hypernuclear low-lying states,
we multiply $\langle {\mathscr F}^{JM}_{k}|$ to the total Schr\"odinger
equation, $\hat{H}|\Psi_{JM}\rangle = E_J|\Psi_{JM}\rangle$, and integrate it over
$\hat{\vec{r}}$ and $\{\vec{r}_N\}$.
This leads to the following coupled-channels equations,
\begin{widetext}
\bsub\begin{eqnarray}
\label{couple1}
&&\left(\frac{d}{dr}-\frac{\kappa-1}{r}\right)
g_{k}(r)+(E_{I_{n}}-E_J) f_{k}(r)
+\sum_{k'}\left[U^{kk'}_V(r)+ U^{kk'}_S(r)\right] f_{k'}(r)
=0, \\
\label{couple2}
&&\left(\frac{d}{dr}+\frac{\kappa+1}{r}\right)
f_{k}(r)-(E_{I_{n}}-2 m_{\Lambda}-E_J)g_{k}(r)
-\sum_{k'}\left[U^{kk'}_V(r) - U^{kk'}_S(r)\right] g_{k'}(r)
= 0,
\end{eqnarray}\esub\end{widetext}
where $\kappa$ is defined as  $\kappa=(-1)^{j+\ell+1/2}(j+1/2)$. With the multipole expansion for the $\delta$ function in coordinate space,
\begin{equation}
\delta(\vec{r}-\vec{r}')=\frac{\delta(r-r')}{rr'}\,
\sum_{\lambda,\mu}Y_{\lambda\mu}(\hat{\vec{r}})Y^*_{\lambda\mu}(\hat{\vec{r}'}),
\end{equation}
the vector and scalar coupling potentials in Eqs.(\ref{couple1}) and (\ref{couple2}) read
\begin{eqnarray}
\label{Coeff_couple1}
U^{kk'}_V(r)
&\equiv&\langle {\mathscr F}^{JM}_{k} |\alpha_V^{N\Lambda}\sum_{i=1}^{A_c}
\delta(\vec{r}-\vec{r}_{Ni})|{\mathscr F}^{JM}_{k'}\rangle \nonumber\\
&&=(-1)^{j'+I+J} \sum_{\lambda}  \langle j\ell  || Y_{\lambda } || j'\ell'  \rangle
\left\{\begin{matrix}
J       &I & j    \\
\lambda &j'  & I' &
\end{matrix}\right\}\nonumber\\
&&\times\alpha_V^{N\Lambda}\rho^{I_{n} I_{n'}}_{\lambda,V}(r),
\end{eqnarray}
and
\begin{eqnarray}
\label{Coeff_couple2}
U^{kk'}_S(r)
&\equiv&\langle {\mathscr F}^{JM}_{k} |\alpha_S^{N\Lambda}\sum_{i=1}^{A_c}\gamma^0_i
\delta(\vec{r}-\vec{r}_{Ni})|{\mathscr F}^{JM}_{k'}
\rangle \nonumber\\
&&=(-1)^{j'+I+J} \sum_{\lambda}  \langle j\ell  || Y_{\lambda } || j'\ell'  \rangle
\left\{\begin{matrix}
J       &I & j    \\
\lambda &j'  & I' &
\end{matrix}\right\}\nonumber\\
&&\times
\alpha_S^{N\Lambda}\rho^{I_{n} I_{n'}}_{\lambda,S}(r),
\end{eqnarray}
where $\rho^{I_{n} I_{n'}}_{\lambda,V}(r)$ and $\rho^{I_{n} I_{n'}}_{\lambda,S}(r)$
are the reduced vector and scalar transition densities 
defined, respectively, as
\bsub\begin{eqnarray}
\label{TD1}
\rho^{I_{n} I_{n'}}_{\lambda, V}(r) &=&
\langle \Phi_{I_n} || \sum\limits_{i=1}^{A_c}
\frac{\delta(r-r_{Ni})}{r_{Ni} r}
Y_\lambda(\hat{\vec{r}}_{Ni})||\Phi_{I_{n'}} \rangle,
\\
\label{TD2}
\rho^{I_{n} I_{n'}}_{\lambda, S}(r) &=&
\langle \Phi_{I_n} || \sum\limits_{i=1}^{A_c} \gamma^0_i
\frac{\delta(r-r_{Ni})}{r_{Ni} r}
Y_\lambda(\hat{\vec{r}}_{Ni})||\Phi_{I_{n'}} \rangle,
\end{eqnarray}
\esub
between the nuclear initial state $I_{n\prime}$ and 
the final state $I_{n}$.
The detailed expression for the transition densities in the
non-relativistic multi-reference DFT framework has been derived
in Ref.~\cite{Yao15} by one of the present authors.
The formalism has also been generalized to the relativistic case within
a multi-reference CDFT (MR-CDFT)
to study a \lq\lq bubble\rq\rq structure in light nuclei~\cite{Yao13PLB,Wu14PRC}. In this work, we extend
this formalism to study low-lying states of hypernuclei.

With the radial wave function ${\mathscr R}_{k}(r)$ in 
the coupled-channels equations (\ref{couple1}) and (\ref{couple2}), 
one can compute the probability $P_{k}$ of the channel $k$ in the 
total wave function $\Psi_{JM}$ as,
\begin{eqnarray}
P_{k}&=&\int r^2 dr\,|{\mathscr R}_{k}(r)|^2 \nonumber\\
&=&\int r^2 dr\left[|f_{k}(r)|^2
+|g_{k}(r)|^2\right].
\end{eqnarray}
The wave function is normalized as $\sum_k P_k=1$.

\subsection{Projected potential energy surface}

In order to apply the formalism presented in the previous subsection, one has to specify
the core states $\Phi_{I_n}$.
A simple choice for this is to construct them as
the projected mean-field states with the intrinsic deformation $\beta$
 \begin{equation}
 \vert \Phi_{IM_I} (\beta)\rangle
 =\hat P^{I}_{M_I K} \hat P^N\hat P^Z\vert \varphi(\beta)\rangle,
\label{proj}
 \end{equation}
where
the particle number projector $\hat P^{N_\tau}$ has the form,
\begin{eqnarray}
  \hat P^{N_\tau}=\frac{1}{2\pi}\int_0^{2\pi}d \varphi_\tau e^{i\varphi_\tau(\hat N_\tau-N_\tau)},\quad (\tau=n,p)
\end{eqnarray}
and the operator $\hat P^{I}_{M_I K} $ is the
three-dimensional angular momentum projection operator given by
\begin{eqnarray}
  \hat P^{I}_{M_IK} =\frac{2I+1}{8\pi^2}\int d\Omega D_{M_I K}^{I*}(\Omega)\hat R(\Omega).
\end{eqnarray}
Here, $\Omega$ represents a set of Euler angles $(\phi, \theta, \psi)$, and the measure
is $d \Omega=d \phi \sin\theta d\theta d\psi$.
$\hat R(\Omega)$ and
$D_{M_I K}^{I}(\Omega)$ are the rotation operator and the Wigner
D-function, respectively~\cite{Edmonds57}.
In Eq. (\ref{proj}),
the wave function $\vert \varphi(\beta)\rangle$ is a Slater determinant of
quasi-particle states with quadrupole deformation $\beta$ generated
with the constrained relativistic mean-field (RMF) calculation.
For simplicity, in this paper we consider only the axial deformation for the nuclear
core and thus the $K$ quantum number is zero in Eq. (\ref{proj}).

In this approach, the hypernuclear states $\Psi_{JM}$ are given for
each deformation $\beta$. This allows one to construct the projected energy
surface for hypernuclei.
That is, one can define the total energy $E_J(\beta)$ by taking
$\langle \Psi_{JM}|\hat{H}|\Psi_{JM}\rangle$ at each deformation $\beta$ and for each spin-parity, $J^\pi$.
Notice that the coupled-channels equations
 (\ref{couple1}) and (\ref{couple2}) are solved at each deformation, and thus the
 effect of core excitations is taken into account in the projected energy surface so obtained.

\subsection{Multi-reference covariant density functional for nuclear core states and transition densities}

One can improve the calculations in the previous subsection
by using $\Phi_{I_n}$
from
the MR-CDFT calculation
in the context of generator coordinate method (GCM)~\cite{Yao10,Yao11,Yao14},
that is,
\begin{equation}
\label{GCM}
 \vert \Phi_{I_n M_I}\rangle
 =\sum_\beta  F^{I}_{n} (\beta)
\hat P^{I}_{M_IK} \hat P^N\hat P^Z\vert \varphi(\beta)\rangle
 \end{equation}
In this wave function, a set of Slater determinants with different quadrupole deformation $\beta$
is superposed according to the idea of GCM.
The weight function  $ F^{I}_{n}(\beta)$  in the
wave function ~(\ref{GCM}) is
determined by requiring that the energy
expectation value is stationary with respect to an arbitrary
variation of $F^{I}_n(\beta)$, which leads to the Hill-Wheeler-Griffin (HWG)
equation~\cite{Griffin57},
\begin{eqnarray}
\label{HWGeq}
  \sum_{\beta'}\left[\mathscr H^{I}_{}(\beta,\beta')-E_{I_{n}} \mathscr N^{I}_{}(\beta,\beta')\right]F^{I}_n(\beta')=0.
\end{eqnarray}
Here,
$\mathscr{N}^{I}_{}(\beta,\beta')=\langle \varphi(\beta)\vert\hat P^I_{00} \hat P^N\hat P^Z\vert \varphi(\beta')\rangle$
and $\mathscr{H}^{I}_{}(\beta,\beta')=\langle \varphi(\beta)\vert\hat H\hat P^I_{00} \hat P^N\hat P^Z\vert \varphi(\beta')\rangle$
are the norm and the energy kernels, respectively. In the calculations,
the energy overlap in the energy kernel is taken to be
the same functional form as in the nuclear mean-field energy, but replacing the densities and
currents with mixed ones, that is, off-diagonal components of the density and current
matrices~\cite{Yao10,Yao11,Yao14}.

Since the projected mean-field states do not form an orthogonal basis and the weights $F^{I}_{n}(\beta)$ in Eq.~(\ref{GCM}) are not orthogonal functions,
it is convenient to construct
a set of orthonormal collective wave functions $g^{I}_{n}$ as~\cite{RS80}
\begin{equation}
\label{wf:g}
g^{I}_{n}(\beta)
= \sum_{\beta'} \big[ \mathcal{N}^{I}_{}\big]^{1/2} 
(\beta,\beta') 
F^{I}_{n}(\beta').
\end{equation}
Notice that
the modulus square of $g^{I}_{n}(\beta)$ does not represent the probability to find the
deformation $\beta$ in the GCM state. For the axial symmetric case, however,
$g^{I}_{n}(\beta)$ provides a good
indication about the dominant configurations in the collective states.

With the GCM wave functions for $|\Phi_{I_n}\rangle$,
one can calculate the reduced vector transition density in Eq.
(\ref{TD1}) as follows~\cite{Yao15},
\begin{eqnarray}
\label{transition_dens}
\rho^{I_{n} I'_{n'}}_{\lambda, V}(r)
&=&(-1)^{I'-I} \dfrac{\hat I^2}{\hat I'}
\sum_{\beta,\beta '}
      F^{I \ast}_{n} (\beta) F^{I^{'}}_{n'} (\beta')\nonumber \\
     && \times\sum_{K\nu}
      \langle I  0 \lambda \nu | I' 0\rangle  \nonumber \\
&&\times
\int  d\hat\br
    Y^\ast_{\lambda\nu}(\hat \br) \langle \varphi(\beta)\vert \hat \rho_{V}(\br)\hat P^{I'}_{0 K}\hat{P}^N \hat{P}^Z |  \varphi(\beta')  \rangle, \nonumber\\
\end{eqnarray}
where the notation $\hat I=\sqrt{2 I +1}$ is introduced for simplicity,
and the vector density operator is defined as follows
\bsub\beqn
\hat\rho_V(\br) = \sum^{A_c}_{i=1} \delta(\br-\br_{Ni}).
\eeqn\esub
The scalar reduced density in Eq. (\ref{TD2}) can also be expressed
in a similar way.

\subsection{Electric quadrupole transition strengths between
hypernuclear states}

The electric quadrupole ($E2$) transition strength from an initial state $|J_i\rangle$ to a final state $|J_f\rangle$ in $\Lambda$ hypernuclei is defined as
\begin{equation}
  B(E2;J_i\rightarrow J_f)= \frac{1}{2J_i+1} \left|\langle J_f ||\hat{Q}_2||J_i \rangle\right|^2.
\end{equation}
Here, the $E2$ operator reads $\displaystyle\hat{Q}_{2\mu}=\sum_{i\in p} r_i^2 Y_{2\mu}(\hat{r}_i)$. Substituting the wave function for the hypernuclear states,
Eq. (\ref{wavefunction}), to this equation, one finds the reduced matrix element to be
\begin{eqnarray}
 \langle J_f ||\hat{Q}_2 ||J_i \rangle
& = &   \sum_{k_i,k_f}  \int dr r^2   {\mathscr R}^\dagger_{k_f} (r) {\mathscr R}_{k_i} (r)
\langle{\mathscr F}^{J_f}_{k_f} \vert\vert \hat{Q}_2 
\vert\vert{\mathscr F}^{J_i}_{k_i}  \rangle.\nonumber \\
\end{eqnarray}
with
\begin{eqnarray}
&&\langle{\mathscr F}^{J_f}_{k_f} \vert\vert \hat{Q}_2 \vert\vert{\mathscr F}^{J_i}_{k_i} \rangle \nonumber \\
&& =  \delta_{j_f j_i}\delta_{\ell_f \ell_i}(-1)^{I_{f}+j_i+J_i}\hat{J_i}\hat{J_f}
 \left\{\begin{matrix}
I_{f} & J_f     &  j_i \\
J_i     & I_{i} &  2
 \end{matrix} \right\} \langle I_{n_f} || \hat{Q}_2 ||I_{n_i} \rangle,\nonumber \\
\end{eqnarray}
(see Eq. (7.1.8) in Ref. \cite{Edmonds57}).
Here,
$\langle I_{n_f} ||\hat{Q}_2||I_{n_i}\rangle $ is the reduced $E2$
transition matrix element between the nuclear core states $\vert I_{f}, n_f\rangle$ and $\vert I_{i}, n_i\rangle$ and it is related to the proton vector transition density in Eq.(\ref{transition_dens}) as
\beq
 \langle I_{n_f} ||\hat{Q}_2||I_{n_i}\rangle
 = \hat I_{i}\int dr  r^{4} \rho^{I_{n_f} I_{n_i}}_{2, V}(r).
 \eeq

\section{Results and discussions}
\label{Sec:Results}

We now numerically solve the coupled-channels equations and discuss low-lying
spectrum of hypernuclei.
The procedure of the calculations and the numerical details are listed
as follows.

{\it (i) Self-consistent  deformation constrained RMF+BCS calculation for the nuclear core states:} This step is to generate a set of deformed states $|\varphi(\beta)\rangle$ with different quadrupole deformation $\beta$. The Dirac equation for nucleons is solved with
the basis of a three-dimensional harmonic oscillator (3DHO) with $N_\mathrm{sh}=10$ major shells. The oscillator length parameter in the 3DHO is chosen as $b_x=b_y=b_z=\sqrt{\hbar/m\omega_0}$, where
$m$ is the nucleon mass and
the oscillator frequency is determined to be 
$\hbar\omega_0=41A_c^{-1/3}$ MeV.
In the most of calculations shown below, we employ
the non-linear point-coupling EDF with the PC-F1~\cite{Buvenich02} set for the
particle-hole channel, although we also use the PC-PK1 set
\cite{PC-PK1}
for $^{20}$Ne
in order to study the parameter set dependence.
In these energy density functionals,
a density independent $\delta$ force
is used
for the particle-particle channel,
supplemented with an energy-dependent cutoff for the pairing 
active space~\cite{Bender00}.

{\it (ii) MR-CDFT calculation for the low-lying states of nuclear 
core:} This step is to obtain the wave functions $\Phi_{I_n}$ for 
the core state $I_n$. The mean-field wave functions are projected 
onto good particle numbers ($N, Z$) and angular momentum $I$, 
which form a set of non-orthogonal basis. The Gauss-Legendre 
quadrature is used for integrals over the Euler angle $\theta$ in 
the calculations of the norm and Hamiltonian kernels.  
For the $^{12}$C nucleus, 
the number of 
mesh points in the interval $[0, \pi]$ for $\theta$ is chosen to be 
14. 
The number of gauge angle $\varphi$
for the particle number projection is
chosen to be 7. 
For $^{20}$Ne and $^{154}$Sm, we use the mesh points of 16 and 9 
for the anglar momentum and the particle number projections, 
respectively. 
The energy and wave function of the core state $I_n$ are determined by solving the HWG equation, Eq. (\ref{HWGeq})
~\cite{Yao10,Yao11,Yao14,Griffin57,RS80}. With the wave functions of nuclear core states, one can calculate the transition densities which are used to determine the coupling potentials in the
coupled-channels equations.

{\it (iii) Coupled-channels calculation for the low-lying states of $\Lambda$ hypernuclei:}
With the coupling potentials so obtained, the coupled-channels equations are solved by expanding
the radial wave function ${\mathscr R}_{j\ell I_{n}}(r)$ on the basis of eigenfunctions of a spherical harmonic oscillator with 18 major shells. From the solutions of the coupled-channels
equations, we construct the spectrum of hypernucleus and calculate the $B(E2)$ transition strengths.

\subsection{Application to $^{13}_{~\Lambda}\mathrm{C}$ }

\subsubsection{Properties of the core states}

\begin{figure}[]
\centering
\includegraphics[width=7cm]{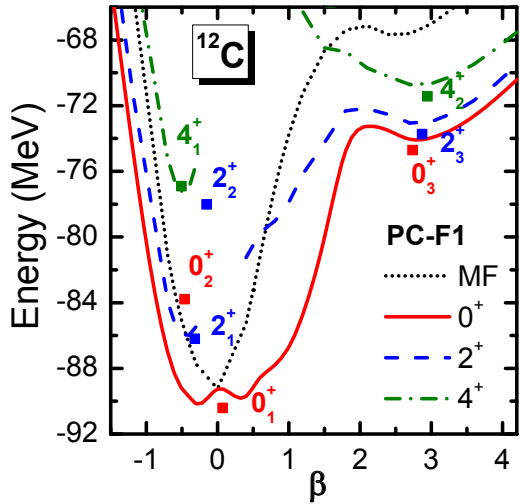}
\caption{(Color online) The energy curve of the mean-field state and the
particle-number and angular momentum projected ($I^\pi=0^+, 2^+$, and $4^+$) states for $^{12}$C  as a function of the intrinsic quadrupole deformation $\beta$.  The filled squares indicate the lowest three GCM solutions for
each $I^\pi$, which are plotted at their average deformation
$\bar\beta\equiv\sum_{\beta} \vert g^{I}_{n} (\beta)\vert^2 \beta$, where $g^{I}_{n} (\beta)$ is
the collective wave functions defined by Eq. (\ref{wf:g}). The results are obtained
with the PC-F1 force.}
\label{fig:PEC-C12}
\end{figure}

We first apply the method to  $^{13}_{~\Lambda}\mathrm{C}$.
Figure~\ref{fig:PEC-C12} shows the energy curves for the core nucleus, $^{12}$C.
The dotted line is obtained in the mean-field approximation, while the other lines
show the projected energy surface for each angular momentum $I$.
The mean-field energy curve exhibits a pronounced minimum at the spherical
configuration
with a steep rise with deformation $\beta$, as expected for a nucleus with large neutron and proton shell gaps. The energy gained from restoration of rotational symmetry increases with deformation $\beta$ and together with
the particle number projection
the location of energy minimum is shifted on the curve. The minimum of the energy curve with $0^+$ is found on the oblate side with $\beta=-0.3$.
Besides, the second minimum appears around $\beta=2.4$ in the mean-field energy curve, which is shifted to $\beta=2.7$ in
the projected energy curve for $I^\pi=0^+$.
It has been shown that the configuration for this minimum has a 3$\alpha$ linear-chain structure~\cite{Arumugam05}.

\begin{figure}[]
\centering
\includegraphics[width=8.5cm]{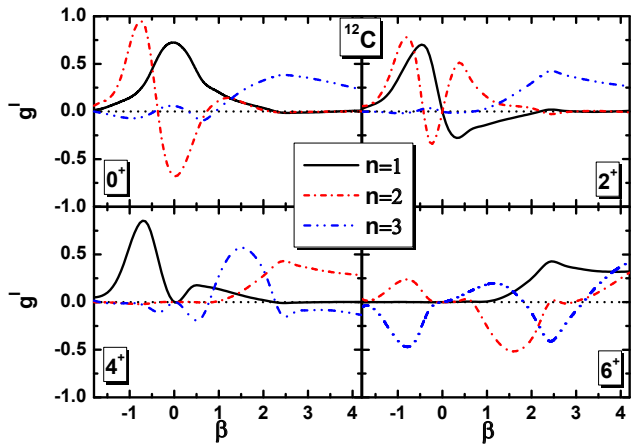}
\caption{(Color online) The collective wave functions $g_n^{I}$ given by Eq.(\ref{wf:g}),
for the first three states in $^{12}$C with spin-parity of $0^+, 2^+, 4^+$ and 6$^+$.}
\label{fig:WF-C12}
\end{figure}

After mixing the projected mean-field configurations, one obtains the energies of
low-lying states (see the filled squares in Fig. \ref{fig:PEC-C12}).
The wave function of these states is displayed in Fig.~\ref{fig:WF-C12}
for the lowest three states with $I=0, 2,4$, and 6.
The ground state of $^{12}$C  is dominated by the spherical configuration.
The collective wave functions and the energy spectrum indicate that there is a coexistence of an anharmonic spherical vibrator and an oblate deformed band at low excitation energies of $^{12}$C. Both structures are not pure and distorted by their strong mixing.
The high-lying $0^+_3, 2^+_3$ and $4^+_2$ states seem to form a rotational band dominated by the $3\alpha$-linear configuration,
in which the collective wave functions
are much extended to a large deformation region.
Similar rotational band corresponding to a $4\alpha$-linear configuration has also been
found in the high-lying states of $^{16}$O~\cite{Yao14-16O}.

\begin{figure*}[htb]
\centering
\includegraphics[width=14cm]{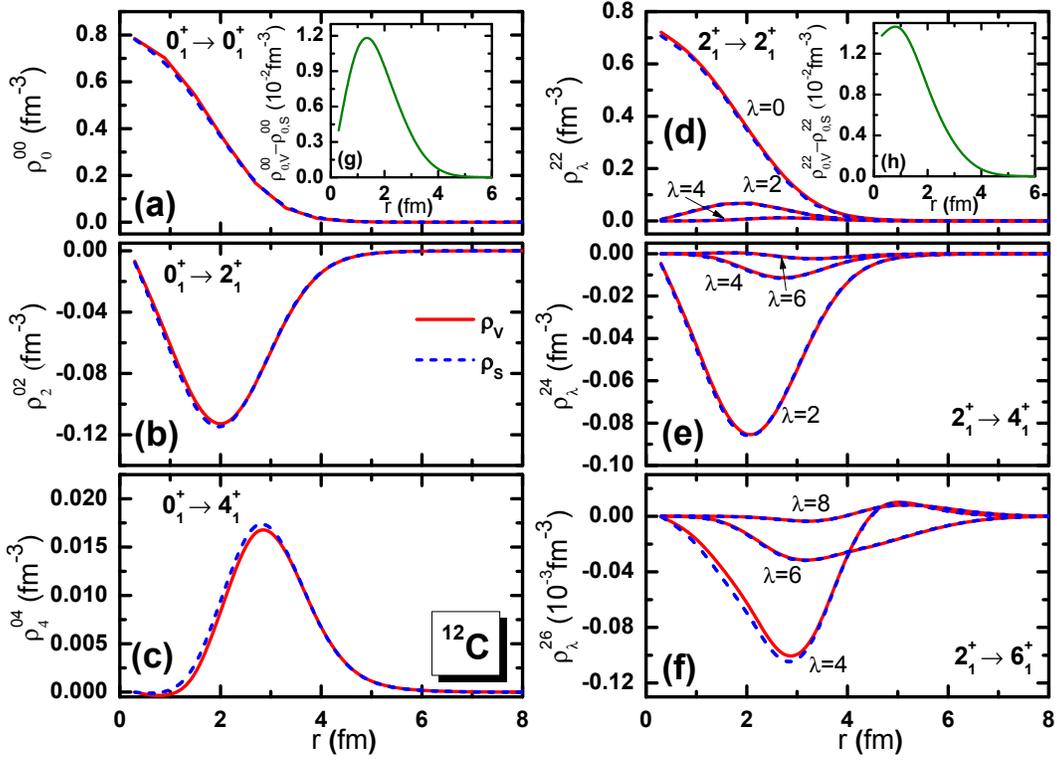}
\caption{(Color online) The vector  transition density, $\rho^{I_{n},I'_{n}}_{\lambda,V}$, given
  by Eq. (\ref{TD1}), and the scalar transition density, $\rho^{I_{n},I'_{n}}_{\lambda,S}$, given by Eq. (\ref{TD2}), in the low-lying states ($n=1$) of $^{12}$C.
These are plotted by the solid and the dashed lines, respectively.
  The insets show the difference of the vector and scalar transition densities.}
\label{fig:Tdensities-C}
\end{figure*}

Figure~\ref{fig:Tdensities-C} shows the vector and scalar transition densities $\rho^{0\lambda}_\lambda$  in the low-lying yrast states ($n=1$) of $^{12}$C, where the multipolarity $\lambda$ is taken as $0, 2$, and $4$. $\rho^{00}_0(r)$ is nothing but the total nucleon density for the $0^+_1$ ground state multiplied by a factor $\sqrt{4\pi}$. It is shown that the transition density
$\rho^{0\lambda}_\lambda$ decreases by one order-of-magnitude as $\lambda$ increases from 0 to 2, and from 2 to 4. Besides, we also plot the transition densities $\rho^{22}_\lambda$ with $\lambda=0, 2$, and $4$, $\rho^{24}_\lambda$ with $\lambda=2, 4$, and $6$, and $\rho^{26}_\lambda$ with $\lambda=4, 6$, and $8$.
Notice that the vector and scalar transition densities are slightly different from one another.

\begin{figure}[]
\centering
\includegraphics[width=8cm]{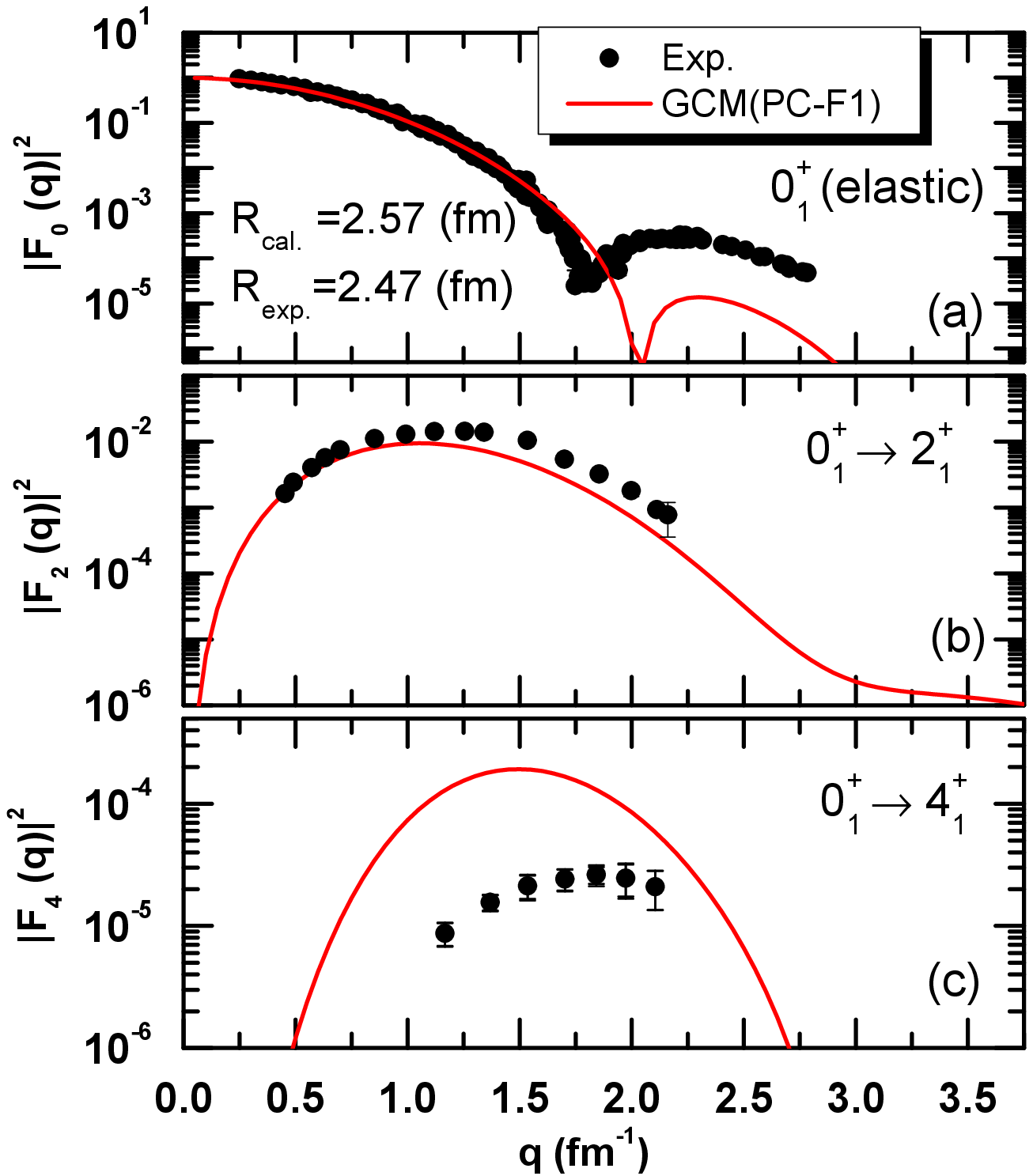}
\caption{(Color online) The charge form factor  for
  the transition from the ground state to
  (a) the ground
state, (b) the excited $2^+_1$ state, and (c) the excited $4^+_1$ state in $^{12}$C calculated with the GCM method
  with the PC-F1 force, in comparison with the available data~\cite{Nakada1971,Cardman80}.
$R$ on the top panel is the root-mean-square charge radius of $^{12}$C.
}
\label{Form:C12}
\end{figure}

\begin{figure}[]
\centering
\includegraphics[width=8cm]{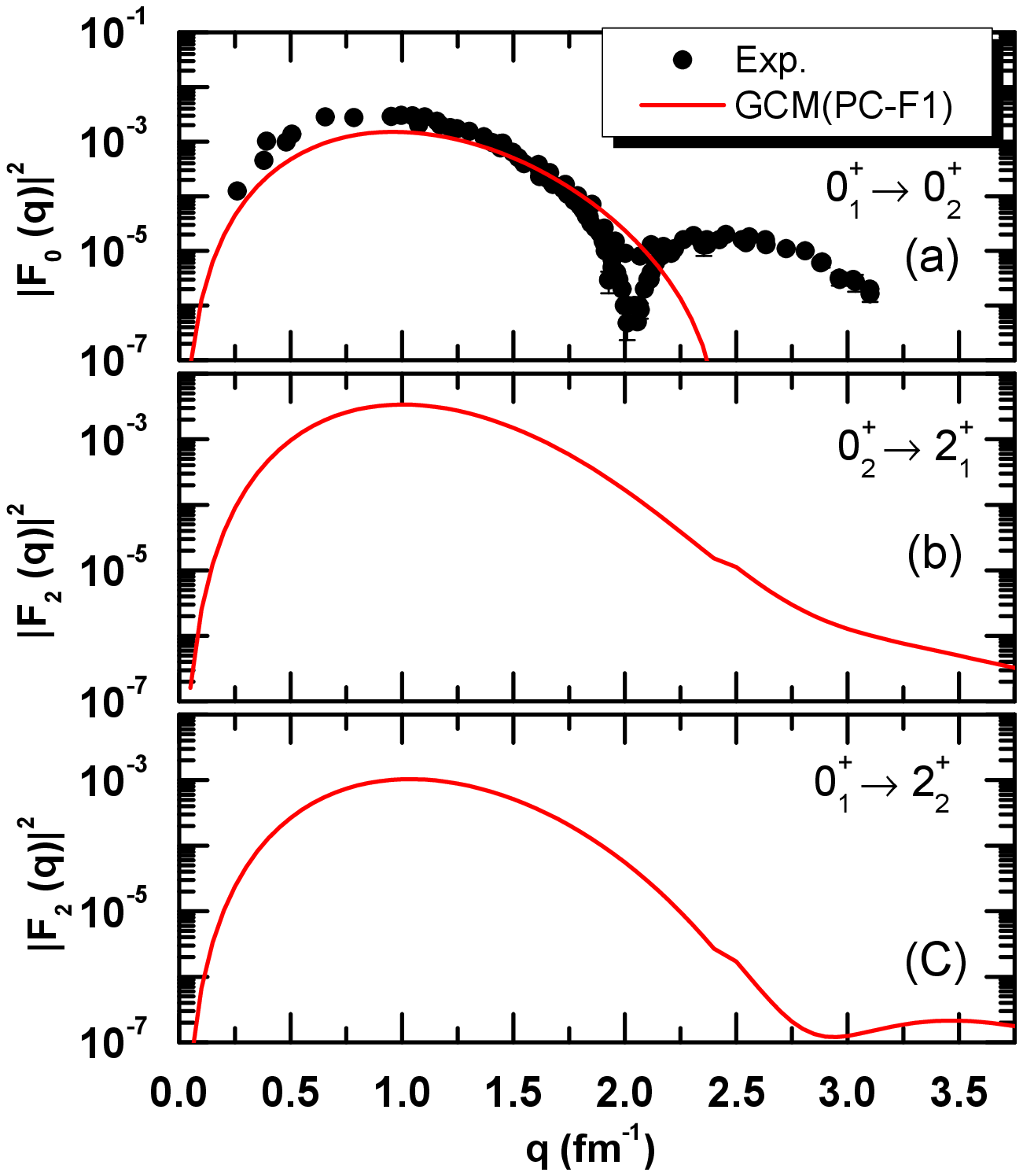}
\caption{(Color online) The charge form factors  between the two bands with $n=1$ and $n=2$ in $^{12}$C calculated with the GCM method with the PC-F1 force, in comparison with the available data from Ref.~\cite{Chernykh07}. }
\label{Form2:C12}
\end{figure}

The convolution of the proton vector transition densities in Fig.
\ref{fig:Tdensities-C} with a
Gaussian form factor for a finite proton size yields the charge transition
densities,
$\rho^{I_n I'_{n'}}_{L, \rm{ch}}(r)$,
which
are related to the form factor $F_L (q)$ for electron scattering
with an angular momentum transfer $L$ by the following
relation~\cite{Yao15},
 \beq
 \label{FF1}
 F_L (q)
 = \dfrac{\sqrt{4\pi}}{Z}
   \int^\infty_0 dr r^2 \rho^{I_n I'_{n'}}_{L, \rm{ch}}(r) j_L (qr),
 \eeq
where $j_L(qr)$ is the spherical Bessel function.
The coefficient $\sqrt{4\pi}/Z$ is chosen so that the elastic
part of the form factor $F_0(q)$ is unity at $q=0$, as in the
experimental data.
A comparison of our results with the data is shown in Fig. ~\ref{Form:C12}.
One can see that the form factors $F_L (q)$ are in rather good agreement with the data except for the underestimation of the elastic form factor after the first minimum, as was found also
in the recent studies for
$^{12}$C~\cite{Fukuoka13} and $^{24}$Mg~\cite{Yao15} based on the Skyrme forces.
This may be because
the spreading of the collective wave function in quadrupole deformation space is somewhat
overestimated in the calculations,
decreasing the weights of the large-$q$ components of the transition density~\cite{Fukuoka13,Yao15}.
In fact, the charge radius of $^{12}$C by the present GCM calculation is 2.57 fm, which is larger
than the empirical value of 2.47 fm~\cite{Angeli04}.

Figure~\ref{Form2:C12} shows the charge form factors for the interband transitions between the two bands with $n=1$ and $n=2$ in $^{12}$C.
Since the $0^+_2$ state
is the Holye state with dilute $3\alpha$ structure, which is beyond the model space of the present calculation, the inelastic form factor $F_0(q)$ corresponding to
the transition from the $0^+_1$ to the $0^+_2$ states
is significantly underestimated in the high-$q$ region beyond the first minimum. It is worthwhile to mention that the calculated electric monopole
transition matrix element $|M(E0: 0^+_2\to 0^+_1)|=4.1$ $e$fm$^2$ and  the charge radius of $0^+_2$ (2.73 fm) are in good agreement with the results ($4.5\pm0.2$  $e$fm$^2$ and $2.73\pm0.02$ fm, respectively) of the recent configuration mixing calculation based on a Skyrme force~\cite{Fukuoka13}.
These values should be compared with the data $|M(E0: 0^+_2\to 0^+_1)|=5.4(2)$ $e$fm$^2$
and the charge radius from other calculations for the Holye state, such as 3.27 fm by the  antisymmetrized molecular dynamics~\cite{Kanada07},
3.38 fm by the fermionic molecular dynamics~\cite{Chernykh07} and  3.83 fm by the alpha-condensation model ~\cite{Funaki03}.

\begin{figure}[]
  \centering
 \includegraphics[width=8.5cm]{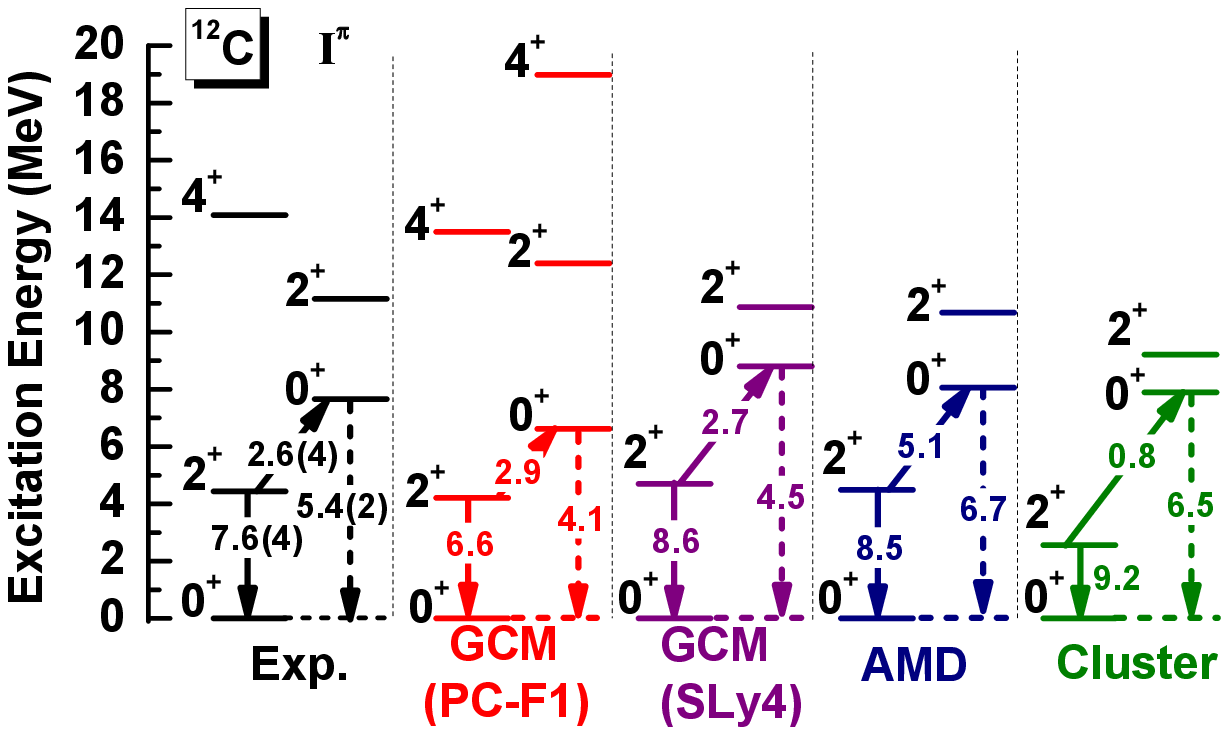}
 \caption{(Color online)
The spectrum of $^{12}$C obtained with several methods, that is,
the GCM with the PC-F1 interaction (the present
calculation), GCM with the Skyrme SLy4 force
~\cite{Fukuoka13}, anti-symmetrized molecular dynamics (AMD)~\cite{Kanada07}
and the alpha cluster model~\cite{Chernykh07}.
The excitation energies are given in units of MeV.
The solid and the dashed arrows are the
quadrupole transition strength $B(E2)$ ($e^2$ fm$^4$) and
the electric monopole transition matrix element $|M(E0)|$ ($e$ fm$^2$), respectively.
The experimental data are taken from Refs. ~\cite{F90,KS05}.
}
  \label{C12Compare}
\end{figure}

The low-lying spectrum of $^{12}$C with MR-CDFT calculation is shown
in  Figure~\ref{C12Compare}, in which the results are compared with
the experiment data \cite{F90,KS05}
as well as with
other model calculations\cite{Fukuoka13,Kanada07,Chernykh07}.
One can see that the low-lying spectrum is reproduced rather well by the present
calculation, although the excitation energies are systematically overestimated.
The electric monopole transition matrix element $|M(E0)|$
and the quadrupole transition strength $B(E2)$ are also
in good agreement with the data and the other model calculations.

\subsubsection{$\Lambda N$ interaction}

\begin{figure}[]
\centering
\includegraphics[width=8.5cm]{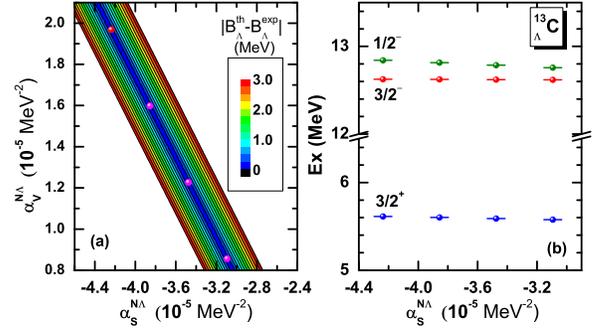}
\caption{(Color online) (a) A contour plot of the absolute value of the difference between the
theoretical and the experimental hyperon binding energies of $^{13}_{~\Lambda}$C hypernucleus
as a function of the coupling strength parameters $\alpha^{N\Lambda}_S$  and $\alpha^{N\Lambda}_V$
in the $N\Lambda$ interaction.
The calculated energies are obtained with the microscopic particle-rotor model.
(b) Energy levels of the $3/2^+$, $3/2^-$ and $1/2^-$ states
in $^{13}_{~\Lambda}$C calculated with the  strength parameters
denoted by the dots in the panel (a). }
\label{Parameter_C12}
\end{figure}

The core states obtained in the previous subsection are used as inputs for the
coupled-channels calculations for the hypernucleus $^{13}_{~\Lambda}$C. To this end,
the parameters $\alpha^{N\Lambda}_S$ and $\alpha^{N\Lambda}_V$ in the $N\Lambda$ interaction,
Eqs.  (\ref{interaction})
and (\ref{interaction2}), are fitted with the microscopic particle-rotor model
to the experimental $\Lambda$ binding energy of
$^{13}_{~\Lambda}$C, that is, $B^{\rm{(exp.)}}_{\Lambda}=11.38\pm 0.05$ MeV~\cite{Hashimoto06}.
Figure \ref{Parameter_C12}(a) shows a contour plot of the absolute value of the
difference between the theoretical and the experimental hyperon binding energies
as a function of $\alpha^{N\Lambda}_S$  and $\alpha^{N\Lambda}_V$.
This is obtained by
including in Eq. (\ref{wavefunction}) the core states up to
$n_{\mathrm{cut}}=2$ and $I_{\mathrm{cut}}=4$.
Obviously,
the two strength parameters cannot be uniquely determined by fitting only to
$B_\Lambda$ and are linearly correlated  as illustrated in Fig.\ref{Parameter_C12}(a).

Taking a few sets of the parameters along the valley with $B_\Lambda^{\rm{th}}=B_\Lambda^{\rm{exp}}$
in Fig. \ref{Parameter_C12}(a),
we calculate the energy of each of the low-lying excited states of $3/2^+$, $3/2^-$ and $1/2^-$ in
$^{13}_{~\Lambda}$C (see Fig.\ref{Parameter_C12}(b)).
One can see that
the excitation energies of $3/2^+$, $3/2^-$ depend on the
choice of the parameters only weakly.  The energy of $1/2^-$ state slightly decreases with the decrease of the absolute value of the
coupling strengths.
For all the sets of the parameters ($\alpha^{N\Lambda}_S, \alpha^{N\Lambda}_V$) in the region of concerned, the energy splitting between
the first $1/2^-$ and $3/2^-$ states is in agreement with the data $152\pm54(\mathrm{stat})\pm36(\mathrm{syst})$ keV~\cite{Ajimura01,Kohri02}. Therefore, as one of the choices, we first fix the value of $\alpha^{N\Lambda}_S$ to be $-4.2377\times10^{-5}$ MeV$^{-2}$,
which is the same value as in the PCY-S2 set~\cite{Tanimura2012}, and
determine
$\alpha_V^{N\Lambda}$ to be $1.969\times10^{-5}$ MeV$^{-2}$.
With this parameter set for the $\Lambda N$ interaction,
the energy splitting between the $1/2^-$ and $3/2^-$ states is
198.9 keV.

\subsubsection{Projected potential energy surface}

 \begin{figure}[]
  \centering
 \includegraphics[width=8.8cm]{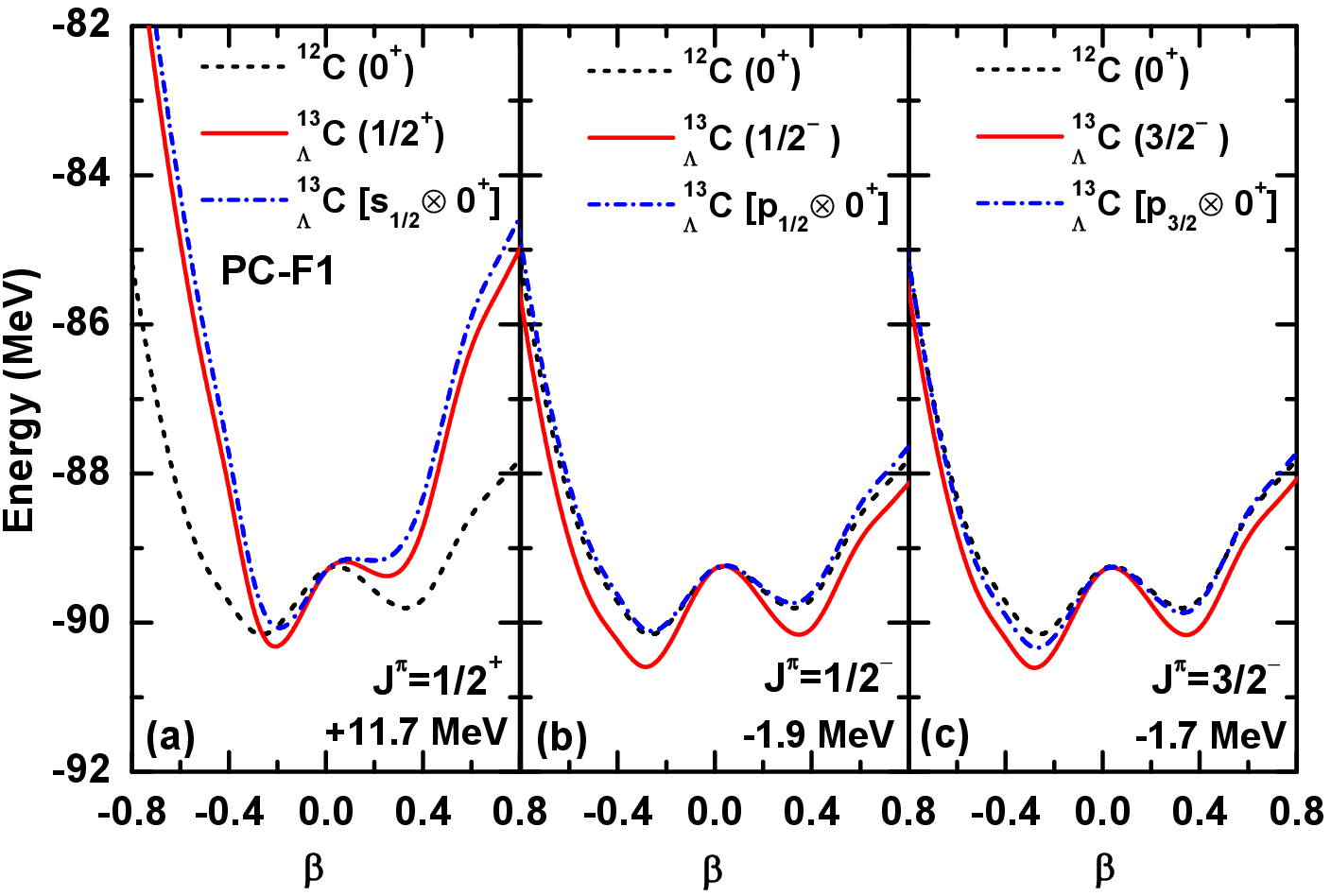}
 \caption{(Color online)The potential energy surfaces of hypernucleus $^{13}_{~\Lambda}$C
obtained with the single-channel calculation (the dot-dashed lines) and the
coupled-channel calculation (the solid lines)
with spin-parity of (a) $J^\pi=1/2^+$, (b) $J^\pi=1/2^-$  and (c) $J^\pi=3/2^-$
as a function of deformation parameter $\beta$.
For comparison, the figure also shows the energy surface for
 the nuclear core with spin-parity of $I^\pi=0^+$ (the dashed lines).
The energy surfaces for hypernuclei are shifted by a constant amount as indicated in each panel.}
  \label{HPEC_C13}
\end{figure}

With the $N\Lambda$ interaction determined in the previous subsection,
let us first investigate the projected potential energy surface for
$^{13}_{~\Lambda}$C.
Figure~\ref{HPEC_C13} shows the resultant energy $E_{J}(\beta)$ for
the  $J^\pi=1/2^+$, $1/2^-$ and $3/2^-$ states
in $^{13}_{~\Lambda}$C  as a function of the deformation $\beta$ of the core nucleus.
These are obtained by solving the coupled-channels equations for each $\beta$.
For comparison, the figure also shows the potential energy curve in the single-channel calculations without
taking into account the core excitations (the dot-dashed lines) as well as those for the
core nucleus (the dashed lines).
The energy surfaces obtained with the coupled-channel calculations are systematically lower than that with  the single-channel calculation due to the additional configuration mixing effect.
One can see that the hypernuclear energy curve with spin-parity of $1/2^+$ has an oblate minimum with $\vert\beta\vert$ significantly
smaller than that of $^{12}$C with $0^+$, indicating a smaller collectivity.
On the other hand, for $J^\pi=1/2^-$, the deformation around the oblate minimum is similar to that of $^{12}$C, but
with a higher barrier at the spherical shape. This leads to a smaller effect of shape mixing
between the prolate and oblate configurations and thus a larger average deformation
in $^{13}_{~\Lambda}$C.
The main component of the $1/2^+$ and $1/2^-$ hypernuclear states are the $\Lambda$ particle in $s_{1/2}$ and $p_{1/2}$ orbits coupled to the ground state of $^{12}$C, respectively. It implies that
a  $\Lambda$ particle in the $s$ ($p$) orbit decreases (increases)
the collectivity of $^{12}$C,
which is consistent with the findings in the recent studies
~\cite{Isaka11,Weixia14}.

\begin{figure*}[]
  \centering
 \includegraphics[width=16cm]{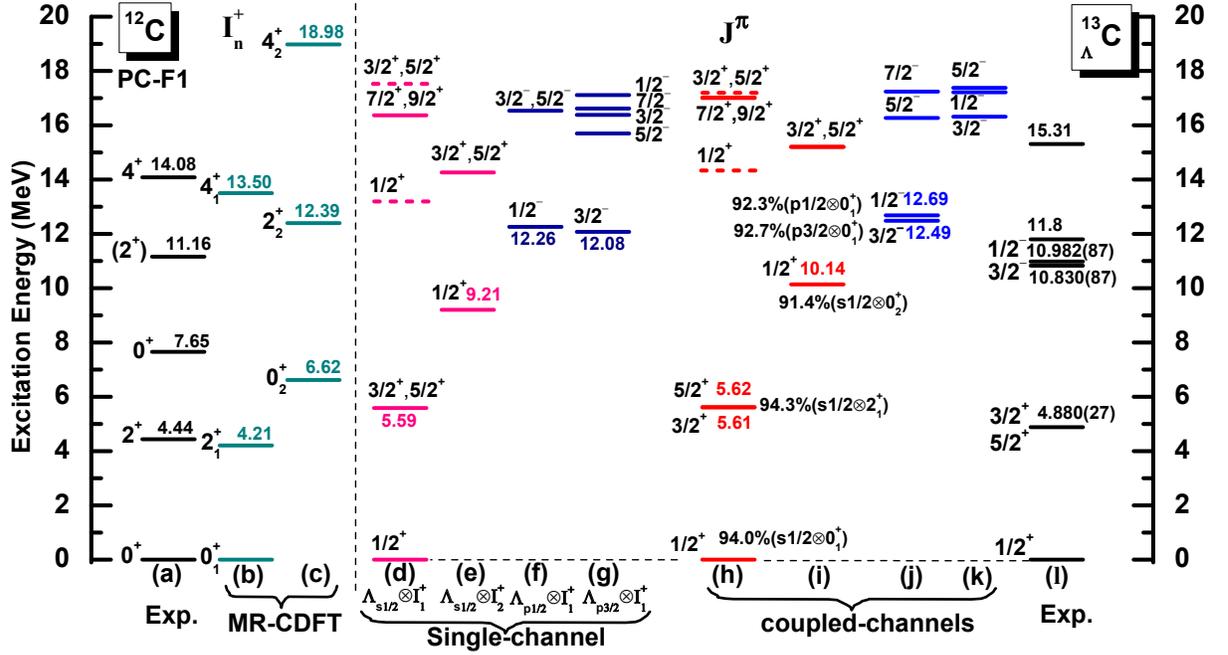}
 \caption{(Color online) The low-energy excitation spectra of
   $^{12}$C ((a)-(c)) and
$^{13}_{~\Lambda}$C ((d)-(l)). For $^{12}$C, the full GCM calculations are compared with the experimental data
taken from Ref.~\cite{NNDC}.  The columns (h) and (i) show the positive-parity states in
$^{13}_{~\Lambda}$C, while the columns (j) and (k) show the negative-parity states.
The experimental data of $^{13}_{~\Lambda}$C are taken from Refs.~\cite{Ajimura01,Kohri02}.
For comparison, the results of single-channel calculation for $^{13}_{~\Lambda}$C with the $\Lambda$ particle in the $s_{1/2}$, $p_{1/2}$ and $p_{3/2}$ orbitals are also plotted in the columns (d), (e),
(f), and (g).}
  \label{C12E}
\end{figure*}

\subsubsection{Single-channel calculations}
\label{sec:12C_single-channel}
Let us now employ the core states described with the GCM method
and discuss the spectrum of $^{13}_{~\Lambda}$C. Before we present the results
of full coupled-channels calculations, we first discuss the results of single-channel
calculations,
restricting the $\Lambda$-hyperon to a specific orbital coupled to a
single core state.
In this case, we take only the diagonal component in the coupling potentials,
Eqs.  (\ref{Coeff_couple1}) and (\ref{Coeff_couple2}), in the coupled-channels
equations.
The coupled-channels equations are then simplified as
\begin{eqnarray}
\label{couple_single}
&&\left(\frac{d}{dr}-\frac{\kappa-1}{r}\right)
g_{k}(r)+(E_{I_n}-E_J) f_{k}(r)  \nonumber \\
&&+\left[U^{kk}_V(r)+ U^{kk}_S(r)\right] f_{k}(r)
=0, \\
&&\left(\frac{d}{dr}+\frac{\kappa+1}{r}\right)
f_{k}(r)-(E_{I_{n}}-2 m_{\Lambda}-E_J)g_{k}(r) \nonumber \\
&&-\left[U^{kk}_V(r) - U^{kk}_S(r)\right] g_{k}(r)
= 0.
\end{eqnarray}

The results for the $\Lambda$ particle in the $s_{1/2}$, $p_{1/2}$ and $p_{3/2}$ orbitals
are shown in
the columns (d),(e),(f) and (g) of Figure~\ref{C12E}.
For comparison, the figure also shows the spectrum of the core nucleus
$^{12}$C
in the columns (a), (b), and (c) (these are actually the same as those
in Fig~\ref{C12Compare}).
A $\Lambda_{\ell j}$ hyperon coupled to the core state with angular momentum $I^+$ produces several hypernuclear states with
$J^\pi$, with the total angular momentum $J$ running from
$J=\vert I - j\vert $ to $J=I + j$, with the parity of $\pi=(-1)^\ell$.
When the $\Lambda$ particle is restricted to
the $s_{1/2}$  orbit,
a doublet states with $(I-1/2)^+$ and $(I+1/2)^+$ are yielded, which
are degenerate in energy for $I>0$.
On the other hand,
a spectrum is more complex for 
the case of $\Lambda$ particle in the $p_{3/2}$ orbital. 
In this case,
the multiplet states
with $J\in [\vert I - j\vert ,(I + j)]$ are
ordered according to the properties of the coupling potential in
Eq.~(\ref{couple_single}).

\begin{figure}[h!]
  \centering
 \includegraphics[width=8cm]{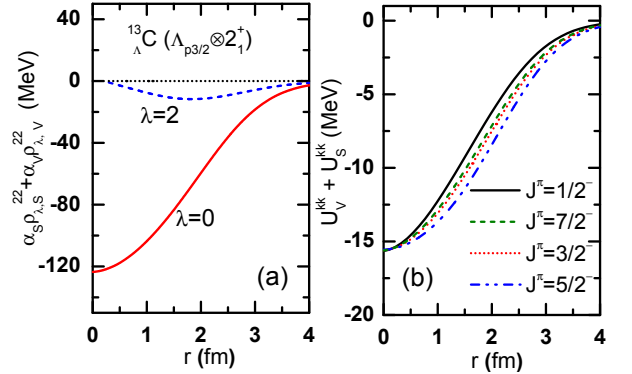}
 \caption{(a) The transition densities $\alpha_V \rho^{2 2}_{\lambda, V}+\alpha_S \rho^{2 2}_{\lambda, S}$ ($\lambda=0$  and $\lambda=2$)
 and (b) the potential $U^{kk}_{V}(r)+U^{kk}_{S}(r)$ in
Eq. (\ref{couple_single}) for the hypernuclear states $J=1/2^-, 3/2^-, 5/2^-, 7/2^-$ with the $\Lambda_{p3/2}\otimes2^+_1$ configuration as a function of the radial coordinate $r$ .}
 \label{rho22_density}
\end{figure}

For instance, in the case of
$\Lambda_{p3/2}\otimes2^+_1$, the multiplets are ordered
as $5/2^-,3/2^-,7/2^-$ and $1/2^-$ (see the column (g) in Fig. ~\ref{C12E}).
In order to understand this, 
we write the coupling potentials for the configuration $k$ as
\begin{eqnarray}
\label{C}
U^{kk}_m
&=&(-1)^{j+I+J} \sum_{\lambda}  \langle j\ell  || Y_{\lambda } || j\ell  \rangle
\left\{\begin{matrix}
J       &I & j    \\
\lambda &j  & I &
\end{matrix}\right\}
\alpha_m^{N\Lambda}\rho^{I_{n} I_{n}}_{\lambda,m} \nonumber\\
&\equiv& C_0 \sum_{\lambda} C_{1\lambda} C_{2\lambda} \alpha^{N\Lambda}_m \rho^{I_{n} I_{n}}_{\lambda, m},
\end{eqnarray}
where the indices $m=S$ and $V$ represent the scalar and vector potentials,
respectively.  The coefficients $C_n~(n=0,1,$ and 2) are defined as $C_0\equiv(-1)^{j+I+J}$,
\begin{eqnarray}
\label{C1}
C_{1\lambda} &\equiv&  \langle j\ell  || Y_{\lambda } || j\ell  \rangle \nonumber \\
    &=& \frac{(-1)^{j+1/2}}{\sqrt{4\pi}} \hat {j}^2 \hat \lambda
    \left(\begin{matrix}
j      &\lambda & j    \\
1/2     & 0      & -1/2 &
\end{matrix}\right) \delta_{\lambda,even},\nonumber \\
\end{eqnarray}
and
\begin{equation}
C_{2\lambda}\equiv\left\{\begin{matrix}
J       &I & j    \\
\lambda &j  & I &
\end{matrix}\right\}.
\label{C2}
\end{equation}
Table~\ref{C_order2} lists the value of each of
the coefficients $C_n$.
The transition densities $\alpha^{N\Lambda}_m \rho^{22}_{\lambda, m}(r)$ and the potential $U^{kk}_{V}(r)+U^{kk}_{S}(r)$ with $k\equiv(j, \ell, I_n)=(\frac{3}{2}, 1, 2_1)$ are displayed in  Fig.~\ref{rho22_density} as a function of radial coordinate $r$. It is seen that the potential $U^{kk}_{V}(r)+U^{kk}_{S}(r)$ becomes gradually deeper
in the order of $J^\pi=1/2^-, 7/2^-, 3/2^-, 5/2^-$, which is consistent with the distribution of energy levels of these multiplets.
It is seen in Table~\ref{C_order2} that
the product $C_0C_{1\lambda}C_{2\lambda}$ is the same
among the multiplets
for $\lambda=0$, and thus
the origin for the energy difference among these four hypernuclear states
is the non-zero $\lambda=2$ term in the potential (\ref{C}).
That is, the splitting of $J^\pi=1/2^-, 7/2^-, 3/2^-$, and $5/2^-$ hypernuclear states is originated from the non-zero transition density $\rho^{22}_2(r)$ (cf. Fig.~\ref{rho22_density}(a)) due to the reorientation effect
(that is, the transition between the same state)
of $2^+_1$ state in the deformed shape of $^{12}$C.
Since
$\alpha^{N\Lambda}_m \rho^{22}_{\lambda, m}(r)$ is negative as shown in Fig.~\ref{rho22_density}(a),
the potential is most attractive for $J^\pi=5/2^-$, which has a positive $C_0C_{12}C_{22}$.

\begin{table}[h!]
\tabcolsep=2pt
\linespread{1.5}
\caption{The coefficients in the potential for the $[\Lambda_{lj}\otimes2^+]^{(J)}$
configurations (see Eqs. (\ref{C1}) and (\ref{C2})).}
 \begin{tabular}{cc|ccc|cc}\hline\hline
$[\Lambda_{lj}\otimes2^+]$& $J^\pi$ & $C_0$ &   $C_{20}$ &$C_0C_{10}C_{20}$ &  $C_{22}$ &$C_0C_{12}C_{22}$ \\ \hline
$[\Lambda_{p3/2}\otimes2^+]$& $1/2^-$ & $~~1.00$   &  $\frac{1}{\sqrt{20}}$     &  $\frac{1}{\sqrt{20\pi}}$    &   $\frac{\sqrt{14}}{20}$    &  $ -\frac{\sqrt{14/\pi}}{20} $   \\
$[\Lambda_{p3/2}\otimes2^+]$& $3/2^-$ & $-1.00$    &   $-\frac{1}{\sqrt{20}} $  &  $\frac{1}{\sqrt{20\pi}}$    &   $~~0.00$                  &  $~~0.00 $   \\
$[\Lambda_{p3/2}\otimes2^+]$& $5/2^-$ & $~~1.00$   &  $\frac{1}{\sqrt{20}} $    &  $\frac{1}{\sqrt{20\pi}}$    &   $ -\frac{\sqrt{14}}{28}$  &  $ \frac{\sqrt{14/\pi}}{28} $  \\
$[\Lambda_{p3/2}\otimes2^+]$& $7/2^-$ & $-1.00$    &   $-\frac{1}{\sqrt{20}} $  &  $\frac{1}{\sqrt{20\pi}}$    &   $ -\frac{\sqrt{14}}{70}$  &  $ -\frac{\sqrt{14/\pi}}{70}$   \\   \hline
$[\Lambda_{p1/2}\otimes2^+]$ & $3/2^-$ & $~~1.00$   &  $\frac{1}{\sqrt{10}}$     &  $\frac{1}{\sqrt{20\pi}}$    &   $~~0.00$    &  $~~0.00$   \\
$[\Lambda_{p1/2}\otimes2^+]$ & $5/2^-$ & $-1.00$    &   $-\frac{1}{\sqrt{10}} $  &  $\frac{1}{\sqrt{20\pi}}$    &   $~~0.00$    &  $~~0.00$  \\ \hline \end{tabular}
\label{C_order2}
\end{table}

For the configuration with $\Lambda$ in $p_{1/2}$ orbital coupled to
the nuclear core $2^+_1$  state,
the resultant
doublet states $ 3/2^-$ and $ 5/2^-$ are degenerate in energy,
since the coefficient $C=C_0C_{1\lambda}C_{2\lambda}$ is not zero only for
$\lambda=0$,
having the same value of $1/\sqrt{20\pi}$
between those two states (see Table I).

 \subsubsection{Coupled-channels calculations}
 \label{sec:12C_couple-channel}

\begin{figure}[]
\centering
\includegraphics[width=8cm]{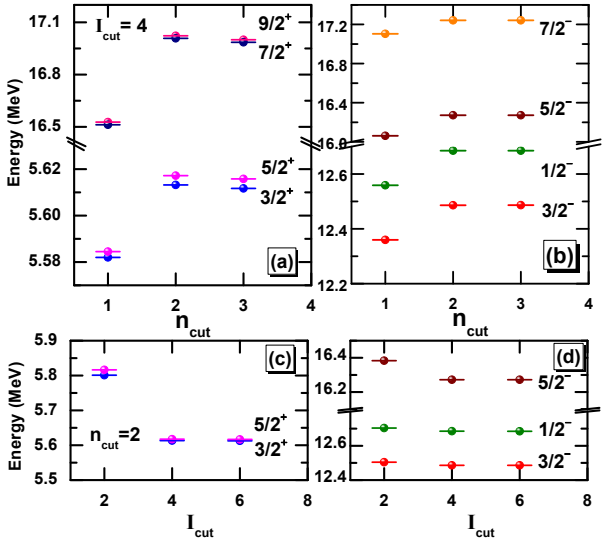}
\caption{(Color online) Excitation energy of
low-lying states in
$^{13}_{~\Lambda}$C as a function of
the cutoff of core states $n$ ((a) and (b)) and
the cutoff of core angular momentum $I$ ((c) and (d))
for the coupled-channels calculations. }
\label{convergence_C12}
\end{figure}

Let us now solve the coupled-channels equations for $^{13}_{~\Lambda}$C.
To this end, we first examine the convergence feature of
the excitation energies
with respect to the cutoff of core states $n_{\rm cut}$ and the
core angular momentum $I_{\rm cut}$.
Fig.\ref{convergence_C12} shows that
$n_{\mathrm{cut}}=2$ and $I_{\mathrm{cut}}=4$ yield a
good convergence for the low-lying excited states, and we use these cut-offs in the
calculations presented below.

The columns (h), (i), (j), and (k) in Figure~\ref{C12E} show
the calculated low-energy excitation spectra of $^{13}_{~\Lambda}$C, in comparison with the corresponding data.
One can see that the low-lying spectra for $^{13}_{~\Lambda}$C are reproduced rather well,
although the excitation energies are slightly overestimated.

In the coupled-channels calculation, the doublets $(5/2^+, 3/2^+)$ and $(9/2^+, 7/2^+)$ in the column (h) mainly consist of the configuration of $\Lambda s_{1/2}\otimes 2^+_1$ and $\Lambda s_{1/2}\otimes 4^+_1$, respectively.
See Tab.~\ref{Component_C12} for the probabilities for the dominant components in each state.
These doublets are degenerate in the single-channel calculation, as already shown in the column (d) in Fig.~\ref{C12E}.
The states of $3/2^+$ and $5/2^+$ are different from each other by 10 keV due to the weak mixing of other configurations. The levels in the column (i) correspond to the configuration of $\Lambda s_{1/2}$ coupled to the second band ($n=2$) in $^{12}$C. These levels share similar features as those in the column (h).

The negative-parity states are shown in the columns (j) and (k) in Fig.~\ref{C12E}.
The energy splitting between the $3/2^-$ and $1/2^-$ states is as small
as 199 keV.
Notice that
in the single-channel calculation the energy difference between the pure configurations of $\Lambda_{p 3/2}\otimes0_1^+$ and $\Lambda_{p 1/2}\otimes0_1^+$ is 180 keV.
That is,
the energy splitting of  $3/2^-$ and  $1/2^-$ states
reflects mainly
the spin-orbit splitting of $\Lambda$ hyperon in the $p_{3/2}$ and $p_{1/2}$ states.
A small splitting between the $3/2^-$ and  $1/2^-$ states has been
shown also in our previous calculation for $^{9}_\Lambda$Be~\cite{Mei14},
although it does not reflect the Lambda spin-orbit splitting
because of a strong mixing between the
$\Lambda_{p_{1/2}}\otimes 0^+$ and the $\Lambda_{p_{3/2}}\otimes 2^+$
configurations in the $1/2^-$ state.

For the second $1/2^-$ and $3/2^-$ states, one can see a large configuration mixing (see Table II).
This is because there are two states whose unperturbed energy in the single-channel
calculations, $E_{\rm 1ch}^{(0)}$,
is close to one another.
These two states are strongly coupled due to the off-diagonal components of the coupling
potentials in the coupled-channels equations.
Notice that, in $^9_\Lambda$Be, this happens already in the
first $1/2^-$ state~\cite{Mei14},
because the reorientation effect discussed in
the previous subsection brings the
$\Lambda_{p_{3/2}}\otimes 2^+$ configuration
close to the $\Lambda_{p_{1/2}}\otimes 0^+$ configuration in energy
due to the prolate nature of the 2$^+$ state of $^8$Be.

According to our calculation, the experimentally observed
level at excitation energy of
11.8 MeV has the spin-parity of $1/2^+$, dominated by the configuration $\Lambda_{s1/2}\otimes 0^+_2$ (c.f. Fig.~\ref{C12E}(e)) or the first radial excitation state of the  configuration $\Lambda_{s1/2}\otimes 0^+_1$ (c.f. Fig.~\ref{C12E}(d)).

\begin{table}[]
\centering
\tabcolsep=6pt
 \caption{The probability $P_{jlI_{n}}$ for the dominant components
in the wave function for low-lying states of $^{13}_{~\Lambda}$C obtained
by the microscopic particle-rotor model.
Only those components which have $P_{jlI_n}$ larger than 0.1 are shown.
$E$ is the energy of each state obtained by solving the coupled-channels
equations, while $E_{\rm 1ch}^{(0)}$ is the unperturbed energy obtained
with the single-channel calculations. The energies are listed in units of MeV. }
 \begin{tabular}{ccccc}
  \hline\hline
     $J^\pi$      &  $E$       &$(l~j)\otimes I^\pi_{n}$      &  $P_{jlI_{n}}$   &  $E^{(0)}_{\rm 1ch}$                             \\ \hline\hline
     $1/2^+_1 $   &  $0.00 $   &  $s_{1/2}\otimes0_1^+$ &  $0.94 $  &  $ 0.00$  \\ \hline
     $3/2^+_1 $   &  $5.61 $   &  $s_{1/2}\otimes2_1^+$ &  $0.94 $  &  $5.59$  \\
     $5/2^+_1 $   &  $5.62 $   &  $s_{1/2}\otimes2_1^+$ &  $0.94 $  &  $5.59$  \\ \hline
     $7/2^+_1 $   &  $17.01 $  &  $s_{1/2}\otimes4_1^+$ &  $0.98 $  &  $ 16.37$  \\
     $9/2^+_1 $   &  $17.02 $  &  $s_{1/2}\otimes4_1^+$ &  $0.98 $  &  $ 16.37$  \\ \hline
     $1/2^+_2 $   &  $10.14$   &  $s_{1/2}\otimes0_2^+$ &  $0.91 $  &  $9.21 $  \\ \hline
     $3/2^+_2 $   &  $15.20$   &  $s_{1/2}\otimes2_2^+$ &  $0.90 $  &  $14.26 $  \\
     $5/2^+_2 $   &  $15.21$   &  $s_{1/2}\otimes2_2^+$ &  $0.90 $  &  $14.26 $ \\ \hline
     $1/2^-_1   $ &  $12.69$   &  $p_{1/2}\otimes0_1^+$ &  $0.92$   &  $12.26$  \\
     $3/2^-_1   $ &  $12.49$   &  $p_{3/2}\otimes0_1^+$ &  $0.93$   &  $12.08$  \\
     $5/2^-_1   $ &  $16.27$   &  $p_{3/2}\otimes2_1^+$ &  $0.82$   &  $15.70$  \\
     $          $ &  $      $  &  $p_{1/2}\otimes2_1^+$ &  $0.17$   &  $16.53$  \\
     $7/2^-_1   $ &  $17.24$   &  $p_{3/2}\otimes2_1^+$ &  $0.97$   &  $16.62$  \\
     $1/2^-_2   $ &  $17.22 $  &  $p_{1/2}\otimes0_1^+$ &  $0.60$   &  $17.36$  \\
     $          $ &  $       $ &  $p_{3/2}\otimes2_1^+$ &  $0.38$   &  $17.11$  \\
     $3/2^-_2   $ &  $16.32 $  &  $p_{3/2}\otimes2_1^+$ &  $0.54 $  &  $ 16.39$  \\
     $          $ &  $       $ &  $p_{1/2}\otimes2_1^+$ &  $0.45 $  &  $ 16.53$  \\
     $5/2^-_2   $ &  $17.38 $  &  $p_{1/2}\otimes2_1^+$ &  $0.80$   &  $16.53 $  \\
     $          $ &  $       $ &  $p_{3/2}\otimes2_1^+$ &  $0.17 $  &  $15.70 $   \\
\hline \hline
 \end{tabular}
    \label{Component_C12}
   \end{table}

 \begin{figure}[]
  \centering
  \includegraphics[width=8.5cm]{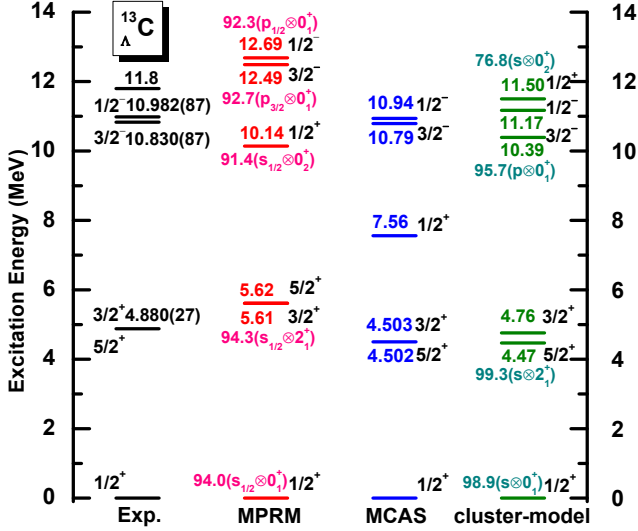}
 \caption{(Color online) A comparison of low-energy excitation spectra of $^{13}_{~\Lambda}$C obtained
with
the present microscopic particle-rotor model (MPRM),
the multi-channel algebraic scattering (MCAS) approach~\cite{Canton2010},
the  $3\alpha+\Lambda$ cluster model~\cite{Hiyama00}, and
the experimental data~\cite{Hashimoto06}.}
  \label{C13:comparison}
\end{figure}

Figure~\ref{C13:comparison} shows a comparison of low-energy excitation spectra of $^{13}_{~\Lambda}$C
obtained with the present microscopic particle-rotor model (MPRM),
the multi-channel algebraic scattering (MCAS) approach~\cite{Canton2010}, and
the $3\alpha+\Lambda$ cluster model~\cite{Hiyama00},
together with the experimental data~\cite{Hashimoto06}.
The basic idea of MCAS approach for $\Lambda$ 
hypernuclei\cite{Canton2010} is similar to the microscopic 
PRM model in a sense that the hypernuclear wave function is given 
by the $\Lambda$ hyperon coupled to the low-lying states of nuclear 
core. In contrast to our full microscopic models, in which all the 
inputs are from the multi-reference CDFT calculation, the MCAS 
approach adopts the experimental data for the energies of nuclear 
core states with an assumption of a pure collective rotational states 
and a phenomenological deformed Woods-Saxon potential for the 
coupling potentials.
In contrast to the MCAS approach and cluster model calculation, 
the ordering of the first degenerate $3/2^+$ and $5/2^+$ 
states are not reproduced in the microscopic PRM calculation, 
since we do not include
a spin-spin interaction in our calculation.
Except for this, the ordering of low-lying states and the structure
of spectrum are
the same between the microscopic PRM and the MCAS approach.
The main components of each state obtained with the microscopic PRM calculation are similar to
those in the cluster model calculation.

Table~\ref{BE2:C} shows the calculated $E2$ transition strengths for low-lying
positive parity states of the
hypernucleus and the corresponding core nucleus. In order to remove the trivial factor due to
the angular momentum coupling for $s_{1/2}$ for the $\Lambda$ particle
and see more clearly the impurity effect of $\Lambda$ particle on
nuclear collectivity, we define the c$B(E2)$ value (that is, the $B(E2)$ value for the
core part) as,
\begin{eqnarray}
&& cB(E2: I_{i}\rightarrow I_{f}) \nonumber \\
&& \equiv \hat{I_{i}}^{-2}\hat{J_f}^{-2} \left\{ \begin{matrix}
I_{f} & J_f     &  j_i \\
J_i     & I_{i} &  2
 \end{matrix} \right\}^{-2}
 B(E2:J_i\rightarrow J_f),
\label{cBE2}
\end{eqnarray}
where $j_i$ is the value for the main channel in the initial state.
The impurity effect of $\Lambda$ particle  can be discussed by comparing the $B(E2)$ values
for the core nucleus and the c$B(E2)$ values for the corresponding hypernucleus.
One can see that the $E2$ transition strength for
$2^+_1\rightarrow 0^+_1$ in $^{12}$C is significantly reduced, by a factor of $\sim14\%$, due to the addition of a $\Lambda$ particle.

\begin{table}[]
\tabcolsep=1.5pt
\caption{The calculated $E2$ transition strengths (in units of $e^2$ fm$^4$) for low-lying
positive parity states of
$^{12}$C and $^{13}_{~\Lambda}$C.
The c$B(E2)$ values are calculated according to Eq. (\ref{cBE2}).
The changes in the $B(E2)$ is indicated with the quantity defined by
$\Delta\equiv (cB(E2)-B(E2; {^{12}{\rm C}}))/B(E2; {^{12}{\rm C}})$.
The value in the parenthesis for $^{12}$C is the experimental data taken
from Ref.~\cite{F90}. }
\label{BE2:C}
\begin{center}
 \begin{tabular}{cc|ccccc}\hline\hline
 \multicolumn{2}{c|}{$^{12}$C}& & \multicolumn{4}{c}{$^{13}_{~\Lambda}$C  }  \\
$I^\pi_i \to I^\pi_f$        & $B(E2)$   &  &  $J^\pi_i \to J^\pi_f$     & $B(E2)$ & $cB(E2)$ &    $\Delta$(\%) \\ \hline
 $2^+_1\rightarrow 0^+_1$    & $ 6.62   $ &  & $3/2^+_1\rightarrow 1/2^+_1$ & $5.68$ & $5.68$     &  $-14.17$      \\
 $                      $    & $ (7.6\pm0.4)  $  &  & $5/2^+_1\rightarrow 1/2^+_1$& $5.68$  & $5.68$     &  $-14.17$     \\
 $4^+_1\rightarrow 2^+_1$    & $14.60 $  &  & $7/2^+_1\rightarrow 3/2^+_1$ & $10.34$ & $11.48$    &  $-21.36$   \\
 $                      $    & $      $  &  & $7/2^+_1\rightarrow 5/2^+_1$ & $1.15$ & $11.49$    &  $-21.35$    \\
 $                      $    & $      $  &  & $9/2^+_1\rightarrow 5/2^+_1$ & $11.48$ & $11.48$    &  $-21.36$    \\
\hline\hline
 \end{tabular}
\end{center}
\end{table}

\subsection{Low-energy spectroscopy of $^{21}_{~\Lambda}\mathrm{Ne}$}

\begin{figure}[]
  \centering
  \includegraphics[width=8cm]{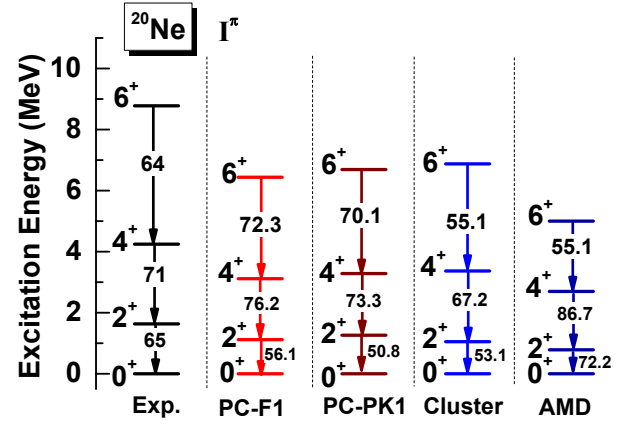}
\caption{(Color online) A comparison
of the yrast rotational states of $^{20}$Ne
obtained with several methods.
The results of the cluster model and the AMD are taken from Refs. ~\cite{Yamada84} and
~\cite{Isaka11}, respectively. }
  \label{Ne20Spetrum}
\end{figure}

We next consider an application to hypernuclei in the $sd$-shell 
region. For this purpose, 
we discuss the $^{21}_{~\Lambda}$Ne hypernucleus.
Since  the $\Lambda$ binding energy in $^{21}_{~\Lambda}$Ne has not yet been measured, we fit the value of coupling strength parameters
$\alpha_S^{N\Lambda}$ and $\alpha_V^{N\Lambda}$
to the $\Lambda$ binding energy estimated with a deformed relativistic mean filed calculation~\cite{Weixia14}. With the PC-F1 and PCY-S1 forces for $NN$ and $N\Lambda$ interactions, respectively, $B_{\Lambda}$ is estimated to be 14.35 MeV for the lowest $\Lambda$ hyperon state. With the same process as in Sec. III A-2,
we obtain a parameter set of $\alpha^{N\Lambda}_S=-4.2377\times10^{-5}$ MeV$^{-2}$ and  $\alpha_V^{N\Lambda}=1.6694\times10^{-5}$ MeV$^{-2}$.

Figure~\ref{Ne20Spetrum} shows the calculated yrast rotational states of $^{20}$Ne.
In order to see the parameter set dependence for the $NN$ interaction, we use both PC-F1
and PC-PK1 parameter sets.
For a comparison, the figure also shows the results of $\alpha+^{16}$O cluster  model~\cite{Yamada84} and the AMD model~\cite{Isaka11}.  One sees that all these models reproduce the rotational character of the yrast states, although they tend to overestimate the moment of inertia.

\begin{figure}[]
  \centering
  \includegraphics[clip=,width=8cm]{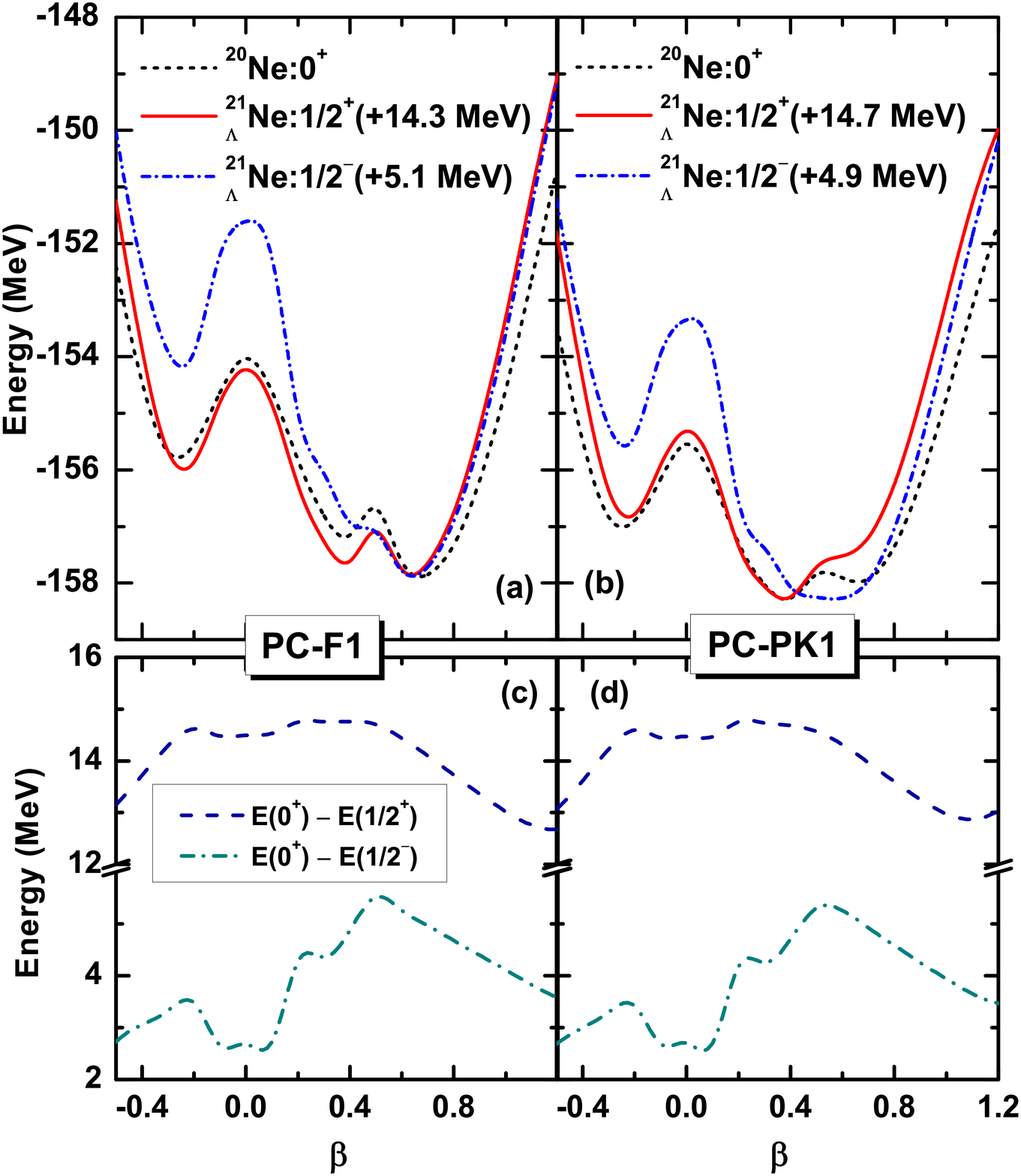}
\caption{(Color online)
Energy curve $E_{J}(\beta)$ for the $J^\pi=1/2^+$ (the solid line) and $ 1/2^-$ (the dot-dashed line) states in  $^{21}_{~\Lambda}$Ne as a function of the deformation $\beta$ of the core nucleus. These are obtained with PC-F1 (the left panel) and PC-PK1 (the right panel) forces.
In order to make a comparison easy, each hypernuclear curve is shifted by a constant value
so that the energy at the absolute minimum coincides with that for $^{20}$Ne with $I^{\pi}=$0$^+$.
  The difference between the energy curve of $^{21}_{~\Lambda}$Ne  and that of
  $^{20}$Ne is shown in the panels (c) and (d) for the PC-F1 and PC-PK1 forces, respectively.}
  \label{Ne20PES}
\end{figure}

Figure~\ref{Ne20PES} shows the obtained energy curve $E_{J}(\beta)$ for the $J^\pi=1/2^+$ and $ 1/2^-$ states in  $^{21}_{~\Lambda}$Ne as a function of the deformation $\beta$ of the core nucleus.
The left and the right panels show the result with PC-F1 and PC-PK1 forces, respectively.
For the latter, we use the same $N\Lambda$ interaction as in the former calculation, even though the
parameters are determined with PC-F1.
We have confirmed that this yields the $B_\Lambda$ value of 
14.33 MeV with PC-PK1, which is
similar to the value with the PC-F1 set, that is, 14.35 MeV.
For PC-F1, the hypernuclear energy curve with spin-parity of $1/2^+$ and $1/2^-$ has a prolate minimum with a smaller $\beta$
than that of $^{20}$Ne with $0^+$.
For PC-PK1, on the other hand,
the value of $\beta$ at the energy minimum remains almost the same for the $1/2^+$ configuration while
that for the $1/2^-$ configuration increases as compared to the deformation for
$^{20}$Ne with $0^+$.
Notice that the energy surface for $1/2^-$ has a higher barrier at the spherical shape than
the barrier for $^{20}$Ne for both the interactions.
This indicates that $^{21}_{~\Lambda}$Ne with $1/2^+$ and $1/2^-$ has a smaller and a larger collectivity than that of $^{20}$Ne, as in $^{13}_{~\Lambda}$C.
The energy differences between the $1/2^+$ state in $^{21}_{~\Lambda}$Ne and the ground state
of $^{20}$Ne, as well as the $1/2^-$ state in $^{21}_{~\Lambda}$Ne and the ground state
of  $^{20}$Ne, are shown in Figs. ~\ref{Ne20PES} (c) and ~\ref{Ne20PES} (d).
Even though PC-F1 and PC-PK1 forces predict somewhat different energy curves,
those energy curves are qualitatively similar to each other, especially when those are plotted
with respect to the energy curve for $^{20}$Ne.

\begin{figure*}[]
  \centering
  \includegraphics[width=16cm]{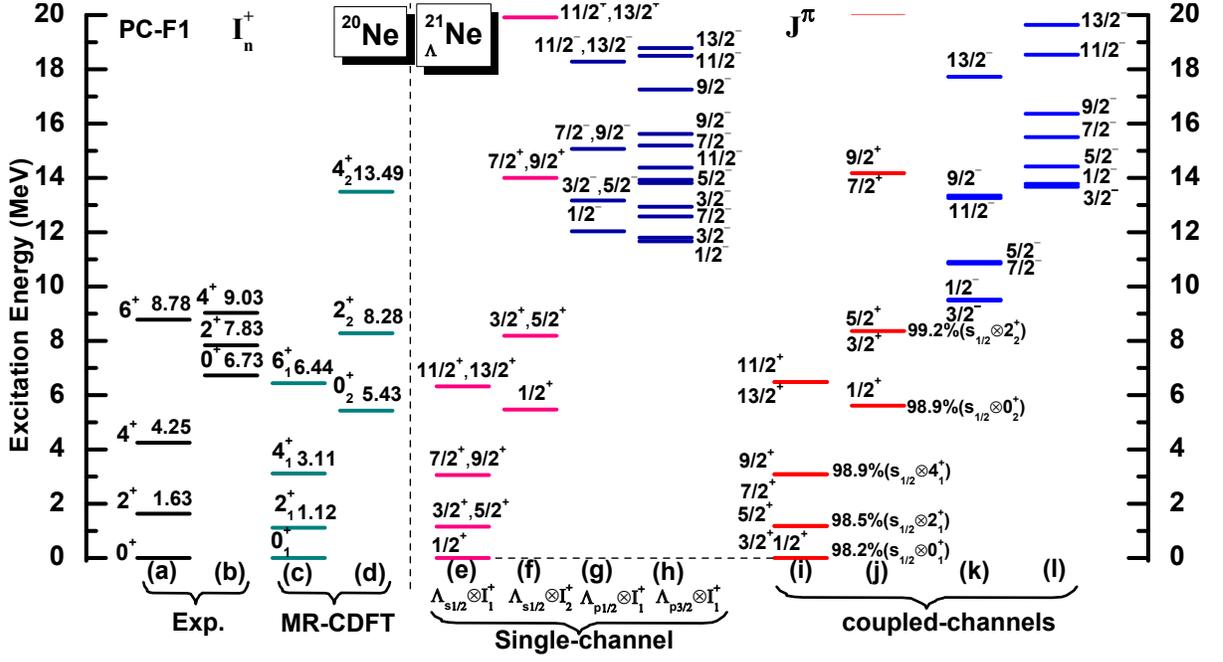}
\caption{(Color online) The low-energy excitation spectra of $^{20}$Ne and $^{21}_{~\Lambda}$Ne
obtained with the microscopic particle-rotor model calculations. }
  \label{Ne21Spetrum}
\end{figure*}

\begin{figure}[]
  \centering
 \includegraphics[width=8cm]{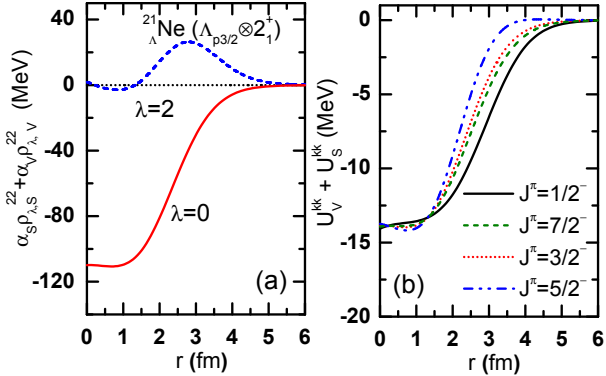}
 \caption{(Color online)Same as Figure~\ref{rho22_density}, but for $^{21}_{~\Lambda}$Ne. }
  \label{rho22_density2}
\end{figure}

\begin{table}[]
\centering
\tabcolsep=3pt
 \caption{Same as Table \ref{Component_C12}, but for $^{21}_{~\Lambda}$Ne hypernucleus with the
PC-F1 and PC-PK1 forces.}
 \begin{tabular}{ccccc|ccc}
  \hline\hline &&  \multicolumn{3}{c|}{PC-F1} &\multicolumn{3}{c}{PC-PK1} \\
   \cline{3-5} \cline{6-8}
     $J^\pi$       &$(l~j)\otimes I^\pi_{n}$   & $E$& $P_{jlI_n}$   &  $E^{(0)}_{\rm 1ch}$ & $E$& $P_{jlI_n}$ & $E^{(0)}_{\rm1ch}$   \\
     \hline\hline
     $1/2^+_1 $    &  $s_{1/2}\otimes0_1^+$ &  $0.00$    & $0.98 $ &$0$       & 0.0 & 0.98 & 0.0 \\  \hline
     $3/2^+_1 $    &  $s_{1/2}\otimes2_1^+$ &  $1.18$    & $0.98 $ &$1.15$    & 1.35& 0.98 & 1.30 \\
     $5/2^+_1 $    &  $s_{1/2}\otimes2_1^+$ &  $1.18$    & $0.98 $ &$1.15$    & 1.35& 0.98 & 1.30 \\  \hline
     $7/2^+_1 $    &  $s_{1/2}\otimes4_1^+$ &  $3.08$    & $0.99 $ &$3.06$    & 3.24& 0.98 & 3.20 \\
     $9/2^+_1 $    &  $s_{1/2}\otimes4_1^+$ &  $3.09$    & $0.99 $ &$3.06$    & 3.24& 0.98 & 3.20 \\  \hline
     $1/2^+_2 $    &  $s_{1/2}\otimes0_2^+$ &  $5.61$    & $0.99 $ &$5.47$    & 5.24& 0.99 & 5.02 \\  \hline
     $3/2^+_2 $    &  $s_{1/2}\otimes2_2^+$ &  $8.36$    & $0.99 $ &$8.19$    & 8.27& 0.99 & 8.04 \\
     $5/2^+_2 $    &  $s_{1/2}\otimes2_2^+$ &  $8.36$    & $0.99 $ &$8.19$    & 8.27& 0.99 & 8.04 \\  \hline
     $1/2^-_1   $  &  $p_{3/2}\otimes2_1^+$ &  $9.52$    & $0.54$  &$11.67$   & 9.55 & 0.52 & 11.75 \\
     $          $  &  $p_{1/2}\otimes0_1^+$ &            & $0.42$  &$12.03$   &      & 0.45 & 11.87 \\
     $3/2^-_1   $  &  $p_{3/2}\otimes0_1^+$ &  $9.48$    & $0.46$  &$11.80$   & 9.50 & 0.48 & 11.64 \\
     $          $  &  $p_{3/2}\otimes2_1^+$ &          & $0.26$  &$12.94$   &      & 0.25 & 13.00 \\
     $          $  &  $p_{1/2}\otimes2_1^+$ &          & $0.24$  &$13.17$   &      & 0.23 & 13.23 \\
     $5/2^-_1   $  &  $p_{1/2}\otimes2_1^+$ &  $10.91$   & $0.46$  &$13.17$   & 10.98& 0.45 & 13.23 \\
     $          $  &  $p_{3/2}\otimes4_1^+$ &         & $0.36$  &$13.93$   &      & 0.37 & 14.00 \\
     $          $  &  $p_{3/2}\otimes2_1^+$ &         & $0.15$  &$13.81$   &      & 0.15 & 13.87 \\
     $7/2^-_1   $  &  $p_{3/2}\otimes2_1^+$ &  $10.85$   & $0.63$  &$12.58$   & 10.92& 0.63  & 12.65 \\
     $          $  &  $p_{1/2}\otimes4_1^+$ &          & $0.19$  &$15.07$   &      & 0.19 &  15.15\\
     $          $  &  $p_{3/2}\otimes4_1^+$ &          & $0.15$  &$15.19$   &      & 0.15 &  15.28\\
\hline \hline
 \end{tabular}
    \label{component_Ne}
   \end{table}

Figure~\ref{Ne21Spetrum} shows the spectra of both $^{20}$Ne and $^{21}_{~\Lambda}$Ne.
To examine the channel-coupling effect on hypernuclear states, we also include the results from single-channel calculations.
The figure only shows the results with PC-F1, since the results with PC-PK1 are similar
(see also Fig. \ref{Ne21Spetrum:comparison} below).
The probability of the main components is summarized in Table~\ref{component_Ne}.
It is shown that the hypernuclear states with positive parity plotted in the
columns (i) and (j) from the full coupled-channels
calculation are close to the results of single-channel calculation shown in the
columns (e) and (f). The analysis of hypernuclear wave functions demonstrates that these states are dominated by the configuration of
$\Lambda_{s1/2}$ coupled to the state $I$ in the first ($n=1$)
and the second ($n=2$) bands in $^{20}$Ne, respectively, with the weight between 98\% and 99\%.
It is seen that the hypernuclear doublet states ($\frac{2I-1}{2}, \frac{2I+1}{2})^\pi$ with configuration $\Lambda_{\ell 1/2}\otimes I^+$ are degenerate, as discussed in Sec.~\ref{sec:12C_single-channel}.  Moreover, it is seen that the spectra of positive-parity states in $^{21}_{~\Lambda}\mathrm{Ne}$ is
close to that of $^{20}$Ne with similar excitation energies to each other. In other words, the presence of a $\Lambda_{s1/2}$ does not change significantly  the low-energy structure of the
core nucleus $^{20}$Ne.

The negative-parity states in $^{21}_{~\Lambda}$Ne
are shown in the columns (k) and (l).
One can see 
that the channel-coupling
effect plays an important role in their excitation energies. 
Moreover, we note that the energy difference between the first $1/2^-$ 
and $3/2^-$ states is less than 40 keV.
Notice that the  $1/2^-$ state is a strong admixture of the configurations 
$\Lambda_{p 1/2}\otimes0^+_1$ and $\Lambda_{p 3/2}\otimes2^+_1$. On the other hand, 
the  $3/2^-$ state is a strong admixture of the configurations 
$\Lambda_{p 3/2}\otimes0^+_1$, $\Lambda_{p 3/2}\otimes2^+_1$ and
$\Lambda_{p 1/2}\otimes2^+_1$.
Therefore, the splitting of the
$1/2^-$ and $3/2^-$ levels in $^{21}_{~\Lambda}\mathrm{Ne}$ does not reflect the strength of $\Lambda$ spin-orbit interaction, which is in marked difference from the case in $^{13}_{~\Lambda}\mathrm{C}$.
From yet another point of view, it is interesting to point out
that a typical rotational band having $L$=1$^-$, 3$^-$, 5$^-, \cdots$
is realized as seen in the column (k) of Fig. \ref{Ne21Spetrum},
apart from the spin of the hyperon.
This group can be charactrized by the $K=0^-$ band based on
the strong coupling between the nuclear rotation and
the hyperon in the $p$-state, and thus
this band manifests a genuinely hypernuclear state with
the [5](90) symmetry which is similar to the [5](50) band
verified in $_\Lambda^9$Be \cite{Motoba83}.
This feature of negative parity states in hypernuclei has already been discussed in Ref. \cite{BIM83}
with the
traditional particle-rotor model with the Elliot SU(3)
model for the core states.

The underlying reason for the difference between $^{13}_{~\Lambda}$C and 
$^{21}_{~\Lambda}$Ne is due to the different properties of the core nuclei. $^{20}$Ne is well-deformed with a much larger transition density $\rho^{02}_2(r)$ than that in $^{12}$C.
Notice also
that the ordering of the $\Lambda_{p 3/2}\otimes2^+_1$ multiplet states is opposite
to that in $^{13}_{~\Lambda}$C, reflecting the fact that the sign of
quadrupole moment is opposite
(that is, prolate deformation for $^{20}$Ne and oblate deformation for $^{12}$C).
In Figure~\ref{rho22_density2}, we plot the transition density
for the $^{21}_{~\Lambda}$Ne hypernucleus. One can see that the
$\lambda$=2 component has the opposite sign as compared to
the transition density for $^{13}_{~\Lambda}$C shown in Fig.
\ref{rho22_density}.
Similar to the $^9_\Lambda$Be case,
these result in several $1/2^-$ and $3/2^-$ states close in energy in the single-channel calculations,
which are strongly mixed in the full coupled-channels treatment.
Similarly to the $1/2^-$ and $3/2^-$ states, the
$5/2^-$ and $7/2^-$ states also show the strong configuration mixing between $\Lambda$-hyperon in
$p_{1/2}$ and $p_{3/2}$ orbits.
We have found that this feature of strong mixing found in $^9_{\Lambda}$Be and $^{21}_{~\Lambda}\mathrm{Ne}$
persists also in heavier systems, such as $^{31}_{~\Lambda}\mathrm{Si}$ and $^{155}_{~~\Lambda}$Sm.

Notice that the $^{20}$Ne nucleus has prominent negative-parity bands originated 
from the $\alpha+^{16}$O
structure. For simplicity, in the present calculations, we have assumed reflection 
symmetry for $^{20}$Ne. The inclusion of these negative
parity states in the coupled-channels calculations is thus 
beyond the scope of the present paper.
It would be an interesting future work to include them and study how the negative 
parity states in $^{21}_{~\Lambda}$Ne
are perturbed.

\begin{figure}[]
  \centering
  \includegraphics[width=8cm]{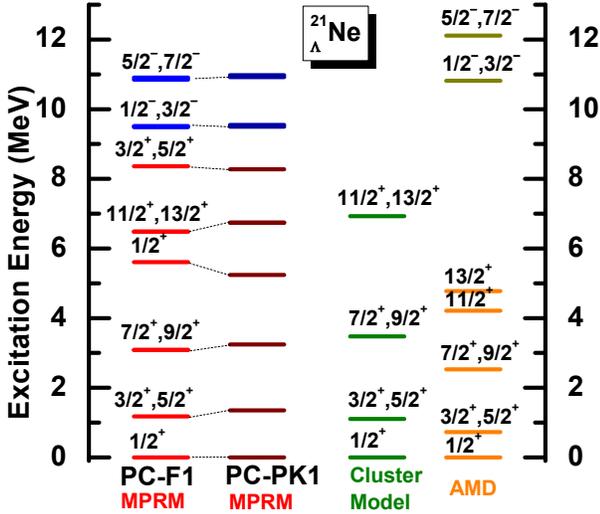}
\caption{(Color online)  A comparison of low-energy excitation spectra of $^{21}_{~\Lambda}$Ne
obtained with the  cluster model~\cite{Yamada84}, the AMD~\cite{Isaka11}, and
the present microscopic particle-rotor model (MPRM) calculations with the PC-F1 and PC-PK1 forces. }
  \label{Ne21Spetrum:comparison}
\end{figure}

Figure~\ref{Ne21Spetrum:comparison} shows a comparison of low-energy excitation spectra of
$^{21}_{~\Lambda}$Ne obtained with
the  cluster model~\cite{Yamada84}, the AMD~\cite{Isaka11}, and
the present microscopic PRM
calculations based on the PC-F1 and PC-PK1 interactions.
The positive-parity band in the microscopic PRM is closer to the result of
cluster model as compared to the result of AMD, which has a
slightly larger moment of inertia.
The negative-parity states are similar to the AMD results but with lower excitation energies,
which might be due to the large channel-coupling
effect taken explicitly into account
in the present work, see Fig.~\ref{Ne21Spetrum} and Table~\ref{component_Ne}.

Table~\ref{BE2:Ne} lists the $E2$ transition strengths for low-lying states of $^{21}_{~\Lambda}$Ne with PC-F1. For comparison, the table also shows the change in $B(E2)$ from
$^{20}$Ne to $^{21}_{~\Lambda}$Ne obtained with the PC-PK1 force,
the  cluster model~\cite{Yamada84}, and the AMD~\cite{Isaka11}.
The $B(E2)$ value decreases by adding a $\Lambda$ hyperon in $s$-orbit in these calculations.
However, the cluster model and AMD model predict more reduction
compared to the microscopic
PRM. A further study would be necessary in order to reconcile this difference.

\begin{table}[]
\tabcolsep=1pt
\caption{The calculated $E2$ transition strengths (in units of $e^2$ fm$^4$) for low-lying states of $^{21}_{~\Lambda}$Ne with the PC-F1 force for the core states.
The results for the change in the $B(E2)$ value from $^{20}$Ne
to $^{21}_{~\Lambda}$Ne is compared with the results
with PC-PK1, the cluster model~\cite{Yamada84} and the AMD~\cite{Isaka11} calculations, where
$\Delta$ is defined in the caption of Table \ref{BE2:C}.
}
\label{BE2:Ne}
\begin{center}
 \begin{tabular}{ccccccc}\hline\hline
  Transition        &  \multicolumn{3}{c}{PC-F1}  &  PC-PK1 & AMD & Cluster  \\
                              \cline{2-4}   \\
 $J^\pi_i \to J^\pi_f$        & $B(E2)$ & $cB(E2)$ &    $\Delta$(\%)&      $\Delta$(\%) & $\Delta$(\%)& $\Delta$(\%)\\ \hline
 $3/2^+_1\rightarrow 1/2^+_1$ & 54.28 & 54.28  & $-3.19$                 & $-7.16$  & $-11.8$&$-23.9$\\
 $5/2^+_1\rightarrow 1/2^+_1$ & 54.28 & 54.28  & $-3.19$                 &  $-7.16$  & $-11.5$&\\
 $7/2^+_1\rightarrow 3/2^+_1$ & 65.90 & 73.22  & $-3.95$                 &$-4.80$  & $-17.8$&$-22.6$\\
 $7/2^+_1\rightarrow 5/2^+_1$ & 7.32  & 73.22  & $-3.95$                 & $-4.80$  & &\\
 $9/2^+_1\rightarrow 5/2^+_1$ & 73.22  & 73.22 & $-3.95$                 & $-4.81$  &$-13.0$ &\\ \hline
 \end{tabular}
\end{center}
\end{table}

We have also applied the microscopic particle-rotor model to study another $sd$- shell hypernucleus, $^{31}_{~\Lambda}$Si.
We have found that
the impurity effect of $\Lambda$ hyperon in $^{31}_{~\Lambda}$Si is qualitatively the same as that in $^{21}_{~\Lambda}$Ne.

\subsection{Low-energy spectroscopy of  $^{155}_{~~\Lambda}\mathrm{Sm}$}

\begin{figure}[]
  \centering
 \includegraphics[width=8.9cm]{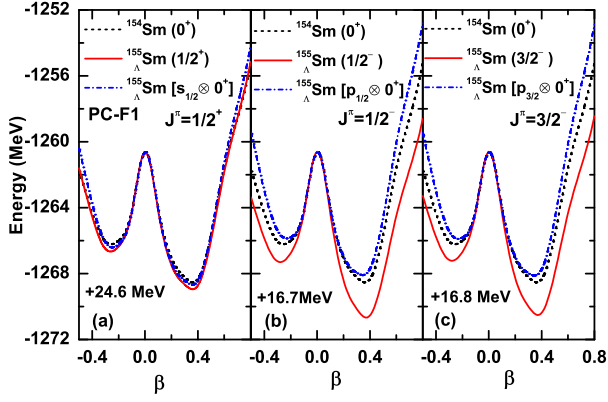}
 \caption{(Color online)Same as Figure~\ref{HPEC_C13}, but for $^{154}$Sm and $^{155}_{~~\Lambda}$Sm. }
  \label{HPEC_Sm155}
\end{figure}

One of the advantages of the microscopic particle-rotor model is
that this method is not limited to light hypernuclei but it can also be applied to medium-heavy
and heavy hypernuclei.
As an example of application to heavy deformed hypernuclei, we next consider $^{154}$Sm
and $^{155}_{~~\Lambda}$Sm.
By fitting to $B_\Lambda$=24.98 MeV estimated with the deformed RMF calculation,
we obtain a parameter set of $\Lambda N$ interaction as
$\alpha^{N\Lambda}_S=-4.2377\times10^{-5}$ MeV$^{-2}$ and $\alpha_V^{N\Lambda}=1.0401\times10^{-5}\mathrm{MeV}^{-2}$.
Figure~\ref{HPEC_Sm155} shows the projected energy curves for $^{154}$Sm and
$^{155}_{~~\Lambda}$Sm obtained with this $N\Lambda$ interaction together with PC-F1 for the
$NN$ interaction.
For the $1/2^+$ state,
the polarization effect of $\Lambda$ particle in $s$-orbit on the properties of $^{154}$Sm is
much smaller than that on $^{12}$C and $^{20}$Ne
due to the large mass number,
although the effect is still large for the negative parity states due to the strong channel
coupling effects.

\begin{figure}[]
  \centering
 \includegraphics[width=9cm]{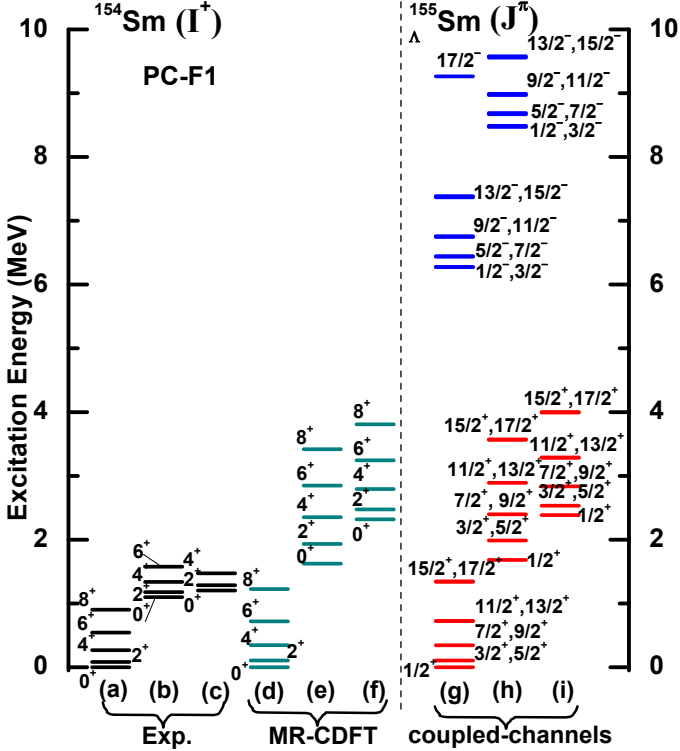}
 \caption{(Color online) The low-energy excitation spectra of
 $^{154}$Sm ((a)-(f)) and  $^{155}_{~~\Lambda}$Sm ((g)-(i)).}
  \label{Sm155E}
\end{figure}

Figure~\ref{Sm155E} shows the calculated low-energy spectrum of  $^{154}$Sm and $^{155}_{~~\Lambda}$Sm  with the PC-F1 force. The ground-state band and the two $\beta$-bands in $^{154}$Sm  are reasonably reproduced, although the band-head energy of the $\beta$-bands are
overestimated.
The low-lying positive-parity states $J^\pi$ in $^{155}_{~~\Lambda}$Sm are dominated by the single-configuration of
$\Lambda_{s_{1/2}}\otimes I^+$ with similar excitation energy as that of the nuclear core
state with $I^+$.
As shown in Tab.~\ref{component_Sm}, the positive-parity states $J^+$, except for $1/2^+$, are nearly two-fold degenerate. These characters are similar to the hypernuclei in the
light-mass region.
On the other hand, one can see
that the negative-parity bands
are well separated from the positive-parity ground band in
$^{155}_{~~\Lambda}$Sm, which is different from the light hypernuclei. It is because the energy scale of the rotational motion is proportional
to $A^{-7/3}$ (see Eq. (1.50) in Ref. \cite{RS80}),
while that of single-$\Lambda$ excitation from $s$ to $p$ orbit is proportional to $A^{-1/3}$. Therefore, with the increase of mass number $A$, the rotational energy spectrum is compressed faster
than the single-$\Lambda$ excitation spectrum. Besides, the low-lying negative-parity states $J^-$  are nearly  two-fold degenerate, even though there are strong configuration-mixing in these states.

Table~\ref{BE2:Sm} presents the $E2$ transition strengths in $^{154}$Sm and $^{155}_{~~\Lambda}$Sm. It is shown that the change in the $B(E2)$ values by adding a $\Lambda$ hyperon in $s$ orbital
is less than $1\%$, which is much smaller than that in the light
hypernuclei studied in this paper.
This is consistent with the small polarization effect of $\Lambda$ particle
discussed in connection to the projected energy surface shown in Fig. ~\ref{HPEC_Sm155}.

\begin{table}[]
\centering
\tabcolsep=6pt
 \caption{Same as Table \ref{Component_C12},but for $^{155}_{~~\Lambda}$Sm hypernuclei.}
 \begin{tabular}{ccccc}
  \hline\hline
     $J^\pi$      &  $E$       &$(l~j)\otimes I^\pi_{n}$      &  $P_{jlI_n}$   &  $E^{(0)}_{\rm 1ch}$                             \\ \hline\hline
     $1/2^+_1 $   &  $0.00$  &  $s_{1/2}\otimes0_1^+$ &  $0.98$  &  $0.00$  \\ \hline
     $3/2^+_1 $   &  $0.11$  &  $s_{1/2}\otimes2_1^+$ &  $0.98$  &  $0.11$  \\
     $5/2^+_1 $   &  $0.11$  &  $s_{1/2}\otimes2_1^+$ &  $0.98$  &  $0.11$  \\ \hline
     $7/2^+_1 $   &  $0.35$  &  $s_{1/2}\otimes4_1^+$ &  $0.98$  &  $0.35$  \\
     $9/2^+_1 $   &  $0.35$  &  $s_{1/2}\otimes4_1^+$ &  $0.98$  &  $0.35$  \\  \hline
     $1/2^+_2 $   &  $1.68$   &  $s_{1/2}\otimes0_2^+$ &  $0.99$  &  $1.58$  \\ \hline
     $3/2^+_2 $   &  $1.99$   &  $s_{1/2}\otimes2_2^+$ &  $0.99$  &  $1.90$  \\
     $5/2^+_2 $   &  $1.99$   &  $s_{1/2}\otimes2_2^+$ &  $0.99$  &  $1.90$ \\  \hline
     $1/2^-_1   $ &  $6.28$   & $p_{3/2}\otimes2_1^+$  &  $0.66$   &  $7.58$  \\
     $          $ &  $$       & $p_{1/2}\otimes0_1^+$  &  $0.32$   &  $8.51$  \\
     $3/2^-_1   $ &  $6.27$   &  $p_{3/2}\otimes0_1^+$ &  $0.35$   &  $8.31$  \\
     $          $ &  $$   &  $p_{3/2}\otimes2_1^+$ &  $0.32$   &  $8.42$  \\
     $          $ &  $$   &  $p_{1/2}\otimes2_1^+$ &  $0.29$   &  $8.61$  \\  \hline
     $5/2^-_1   $ &  $6.44$   &  $p_{3/2}\otimes4_1^+$ &  $0.53$   &  $8.08$  \\
     $          $ &  $ $  & $p_{1/2}\otimes2_1^+$  &  $0.33$   &  $8.61$  \\
     $          $ &  $ $  &  $p_{3/2}\otimes2_1^+$ &  $0.11$   &  $8.96$  \\
     $7/2^-_1   $ &  $6.44$   &  $p_{3/2}\otimes2_1^+$ &  $0.47$   &  $8.19$  \\
     $          $ &  $$   &  $p_{1/2}\otimes4_1^+$ &  $0.28$   &  $8.86$  \\
     $          $ &  $$   &  $p_{3/2}\otimes4_1^+$ &  $0.22$   &  $8.89$  \\
\hline \hline
 \end{tabular}
    \label{component_Sm}
   \end{table}

\begin{table}[]
\tabcolsep=1.2pt
\caption{Same as Table \ref{BE2:C}, but for $^{154}$Sm and $^{155}_{~~\Lambda}$Sm.
The experimental data for $^{154}$Sm, shown in the parenthesis, is taken
from Ref. \cite{lbl}. }
\label{BE2:Sm}
\begin{center}
 \begin{tabular}{cc|ccccc}\hline\hline
 \multicolumn{2}{c|}{$^{154}$Sm }& & \multicolumn{4}{c}{$^{155}_{\Lambda}$Sm   }  \\
$I^\pi_i \to I^\pi_f$       & $B(E2)$  &&  $J^\pi_i \to J^\pi_f$     & $B(E2)$ & $cB(E2)$ & $\Delta$(\%) \\ \hline
$2^+_1\rightarrow 0^+_1$    & $9358.49$  && $3/2^+_1\rightarrow 1/2^+_1$& $9284.69$ & $9284.69$ & $-0.79$    \\
$                      $    & $(8720\pm100)$&& $5/2^+_1\rightarrow 1/2^+_1$&$9284.23$ &$9284.23$ &$-0.79$    \\
$4^+_1\rightarrow 2^+_1$    & $13512.18$ && $7/2^+_1\rightarrow 3/2^+_1$& $12081.04$ &$13423.38$&$-0.66$    \\
$                      $    & $      $   && $7/2^+_1\rightarrow 5/2^+_1$& $1342.25$ & 13422.54& $-0.66$   \\
$                      $    & $      $   && $9/2^+_1\rightarrow 5/2^+_1$& $13422.25$ & $13422.25$& $-0.67$   \\
\hline\hline
 \end{tabular}
\end{center}
\end{table}

\section{Summary}
\label{Sec:Summary}

 We have presented the detailed formalism of the microscopic particle rotor model  based on a covariant density functional theory  for the low-lying states of single-$\Lambda$ hypernuclei.
In this formalism, the wave functions for hypernuclei
have been constructed by coupling the $\Lambda$ hyperon to the low-lying states of the core nucleus.
The radial wave functions are obtained by
solving the corresponding coupled-channel equations, in which the coupling potentials
are provided in terms of the transition densities of the
nuclear core states. For simplicity, in this paper
we have adopted only the leading-order four-fermion coupling terms of scalar and vector types for the $\Lambda N$ effective interaction.
Applying this method to $^{13}_{~\Lambda}$C,
we have reproduced reasonably well the experimental energy spectrum
of this hypernucleus.
We have applied this method also to
$^{21}_{~\Lambda}$Ne,
and $^{155}_{~~\Lambda}$Sm,
and have achieved a good agreement with other model studies both for
the excitation energies and the compositions of wave functions.
We mention that our method is the only one which
can be applied to such heavy hypernuclei.
We have found that the $NN$ interaction with the PC-F1 and PC-PK1 sets
lead to similar
hypernuclei spectra to each other.
For all the hypernuclei, the low-lying excited states with positive parity $J^+$, except for $1/2^+$, are nearly two-fold degenerate and dominated by the single-configuration of
$\Lambda_{s_{1/2}}\otimes I^+$, where $I^+$ is the spin-parity of the nuclear core states.
In contrast, in general there are large configuration mixing in the negative-parity states.
We have, however, found an exception for this, that is,
for the first $3/2^-$ and $1/2^-$ states in $^{13}_{~\Lambda}$C
the effect of configuration mixing is rather small, and thus the energy splitting of these states
reflects the spin-orbit splitting of $\Lambda$ hyperon in the $p_{3/2}$ and $p_{1/2}$ states.
Concerning the electromagnetic transitions, we have found that for all the systems
the $B(E2)$ value from the first $2^+$ to the ground states in the core nuclei
is reduced by adding a $\Lambda$ particle in the positive-parity states.
The reduction factor is about
$14\%$ for $^{13}_{~\Lambda}$C,
$3.2\%$ for $^{21}_{~\Lambda}$Ne,
and $0.79\%$ for $^{155}_{~~\Lambda}$Sm, and thus the reduction factor is
larger for the oblate hypernuclei.
For $^{21}_{~\Lambda}$Ne and $^{31}_{~\Lambda}$Si,
a slightly larger impurity effect was found with the PC-PK1 force
as compared to the PC-F1 force.

New measurements of $\gamma$-ray spectroscopy of hypernuclei will soon
start at the new generation experimental facilities such as J-PARC.
It would be interesting if the low-lying spectra predicted in this
paper are confirmed in near future.

\section*{Acknowledgments}
This work was supported in part by the Tohoku University Focused Research Project \lq\lq Understanding the origins for matters in universe\rq\rq, JSPS KAKENHI Grant Number 
26400263, the National Natural Science Foundation of China
under Grant Nos. 11305134, 11105111, and the Fundamental Research Funds for the Central
University (XDJK2013C028).


\begin{thebibliography}{99}{}

\bibitem{Hashimoto06} O. Hashimoto and H. Tamura, Prog. Part. Nucl. Phys.
\textbf{57}, 564 (2006).

\bibitem{Tamura09} H. Tamura, Int. J. Mod. Phys. A \textbf{24}, 2101 (2009).

 \bibitem{Motoba83}T. Motoba, H. Band\={o}, and K. Ikeda,
Prog. Theor. Phys. \textbf{70}, 189 (1983).

\bibitem{Hiyama99} E. Hiyama, M. Kamimura, K. Miyazaki, and T. Motoba,
Phys. Rev. C \textbf{59}, 2351 (1999).

\bibitem{Bando90}  H. Bando, T. Motoba and J. \v{Z}ofka,
Int. J. Mod. Phys. \textbf{A 5}, 4021 (1990).

\bibitem{Hiyama03} E. Hiyama, Y. Kino, and M. Kamimura,
Prog. Part. Nucl. Phys. \textbf{51}, 223 (2003).

\bibitem{Cravo02} E. Cravo, A. C. Fonseca, Y. Koike,
Phys. Rev. C \textbf{66}, 014001 (2002).

\bibitem{Suslov04} V. M. Suslov, I. Filikhin, and B. Vlahovic,
J. Phys. G: Nucl. Part. Phys. \textbf{30}, 513 (2004).

\bibitem{Shoeb09}M. Shoeb and Sonika, Phys. Rev. C \textbf{79}, 054321 (2009).

\bibitem{Dalitz78} R. H. Dalitz and A. Gal, Ann. Phys. (N.Y.)
\textbf{116}, 167 (1978).

\bibitem{Gal71}A. Gal, J.M. Soper, and R.H. Dalitz,
Ann. Phys. (N.Y.) {\bf 63}, 53 (1971).

\bibitem{Millener}D. J. Millener, Nucl. Phys. {\bf A804}, 84 (2008);
{\bf A914}, 109 (2013).

\bibitem{abinitio}
R. Wirth, D. Gazda, P. Navratil, A. Calic, J. Langhammer, and
R. Roth, Phys. Rev. Lett. \textbf{113}, 192502 (2014).

\bibitem{Isaka11} M. Isaka, M. Kimura, A. Dot\'e and A. Ohnishi,
Phys. Rev. C \textbf{83}, 044323 (2011).

\bibitem{Isaka11-2}
M. Isaka, M. Kimura, A. Dot\'e and A. Ohnishi,
Phys. Rev. C \textbf{83}, 054304 (2011).

\bibitem{Isaka12}
M. Isaka, H. Homma, M. Kimura, A. Dot\'e and A. Ohnishi,
Phys. Rev. C \textbf{85}, 034303 (2012).

\bibitem{Isaka13}
M. Isaka, M. Kimura, A. Dot\'e and A. Ohnishi,
Phys. Rev. C \textbf{87}, 021304(R) (2013).

\bibitem{Zhou07} X. R. Zhou ,
H.-J. Schulze, H. Sagawa, C. X. Wu, and E.-G. Zhao,
Phys. Rev. C \textbf{76}, 034312 (2007).

\bibitem{Win08} M. T. Win and K. Hagino, Phys. Rev. C \textbf{78}, 054311
(2008).

\bibitem{Schulze10} H.-J. Schulze, M. T. Win, K. Hagino, and H. S. Sagawa,
Prog. Theo. Phys. \textbf{123}, 569 (2010).

\bibitem{Win11} Myaing Thi Win, K. Hagino, and T. Koike,
Phys. Rev. C \textbf{83}, 014301 (2011).

\bibitem{Lu11} B.-N. Lu, E.-G. Zhao, and S.-G. Zhou,
Phys. Rev. C \textbf{84}, 014328 (2011).

\bibitem{Weixia14} W. X. Xue, J. M. Yao, K. Hagino, Z. P. Li, H. Mei, and Y. Tanimura,
Phys. Rev. C \textbf{91}, 024327 (2015).

\bibitem{Li13} A. Li, E. Hiyama, X.-R. Zhou, and H. Sagawa,
Phys. Rev. C \textbf{87}, 014333 (2013).

\bibitem{Lu14} B.-N. Lu, E. Hiyama, H. Sagawa, and S.-G. Zhou,
Phys. Rev. C \textbf{89}, 044307 (2014).

\bibitem{HY14}K. Hagino and J.M. Yao,
arXiv:1410.7531.

\bibitem{Mei14} H. Mei, K. Hagino, J.M. Yao, and T. Motoba,
 Phys. Rev. C \textbf{90}, 06430 (2014).

\bibitem{Tanimura2012}Y. Tanimura and K. Hagino,
Phys. Rev. C \textbf{85}, 014306 (2012).

\bibitem{Buvenich02} T. Burvenich, D. G. Madland, J. A. Maruhn, and P.-G. Reinhard,
Phys. Rev. C \textbf{65}, 044308 (2002).

\bibitem{Yao15} J. M. Yao, M. Bender, and P.-H. Heenen,
Phys. Rev. C {\bf 91}, 024301 (2015).

\bibitem{Yao13PLB} J. M. Yao, H. Mei, Z. P. Li,
Phys. Lett. B \textbf{723}, 459 (2013).

\bibitem{Wu14PRC} X. Y. Wu, J. M. Yao, and Z. P. Li,
Phys. Rev. C \textbf{89}, 017304 (2014).

\bibitem{Edmonds57} A. Edmonds, {\it Angular Momentum in Quantum Mechanics}
(Princeton University Press, Princeton, NJ, 1957).

\bibitem{Yao10} J. M. Yao, J. Meng, P. Ring, and D. Vretenar,
Phys. Rev. C \textbf{81}, 044311(2010).

\bibitem{Yao11}
J. M. Yao, H. Mei, H. Chen, J. Meng, P. Ring, and
D. Vretenar, Phys. Rev. C \textbf{83}, 014308 (2011).

\bibitem{Yao14}
J. M. Yao, K. Hagino, Z. P. Li, J. Meng, and P. Ring,
Phys. Rev. C  \textbf{89}, 054306 (2014).

\bibitem{Griffin57} J. J. Griffin, J. A. Wheeler, J. J. Griffin, and J. A.
Wheeler, Phys. Rev. {\bf 108}, 311 (1957).

\bibitem{RS80} P. Ring and P. Schuck,
 {\it The Nuclear Many Body Problem}
(Springer-Verlag, New York, 1980).

\bibitem{PC-PK1}
P.W. Zhao, Z.P. Li, J.M. Yao,
and J. Meng, Phys. Rev. C{\bf 82}, 054319 (2010).

\bibitem{Bender00} M. Bender, K. Rutz, P.-G. Reinhard, and J. A. Maruhn,
 Eur. Phys. J. A 8, \textbf{59} (2000).

\bibitem{Arumugam05}  P. Arumugam, B. K. Sharma, S. K. Patra, and R. K. Gupta,
 Phys. Rev. C \textbf{71}, 064308 (2005).

\bibitem{Yao14-16O}  J. M. Yao, N. Itagaki, J. Meng,
Phys. Rev. C \textbf{90}, 054307 (2014).

\bibitem{Fukuoka13} Y. Fukuoka, S. Shinohara, Y. Funaki, T. Nakatsukasa, and K. Yabana,
 Phys. Rev. C \textbf{88}, 014321 (2013).

\bibitem{Angeli04} I. Angeli,
At. Data Nucl. Data Tables \textbf{87}, 185 (2004).

 \bibitem{Kanada07} Y. Kanada-En'yo,
 Prog. Theor. Phys. \textbf{117}, 655 (2007).

\bibitem{Chernykh07} M. Chernykh, H. Feldmeier, T. Neff, P. von Neumann-Cosel, and A. Richter,
 Phys. Rev. Lett. \textbf{98}, 032501 (2007).

\bibitem{Funaki03} Y. Funaki, A. Tohsaki, H. Horiuchi, P. Schuck, and G. R\"{o}pke.
Phys. Rev. C \textbf{67}, 051306(R) (2003).

\bibitem{F90} F. Ajzenberg-Selove,
Nucl. Phys. {\bf A506}, 1 (1990).

\bibitem{KS05}
T. Kibedi and R.H. Spear, At. Data and Nucl. Data Tables,
{\bf 89}, 77 (2005).

\bibitem{Nakada1971} A. Nakada, Y. Torizuka and Y. Horikawa,
 Phys. Rev. Lett. \textbf{27}, 745 (1971).

 \bibitem{Cardman80} L. S. Cardman, , J.W. Lightbody, S. Penner, W.P. Trower, S.F. Williamson,
 Phys. Lett. \textbf{B 91}, 203 (1980).

\bibitem{NNDC} National Nuclear Data Center (NNDC), http://www.nndc.bnl.gov/.

\bibitem{Ajimura01} S. Ajimura, H. Hayakawa, T. Kishimoto, H. Kohri, K. Matsuoka, S. Minami, T. Mori, K. Morikubo, E. Saji, A. Sakaguchi {\it et al}
Phys, Rev. Lett. {\bf 86},4255 (2001).

\bibitem{Kohri02}H. Kohri, S. Ajimura, H. Hayakawa, T. Kishimoto, K. Matsuoka,
S. Minami, Y. S. Miyake, T. Mori, K. Morikubo, E. Saji {\it et al}
Phys.Rev.C {\bf 65},034607 (2002).

\bibitem{Canton2010} L. Canton, K. Amos, S. Karataglidis, J. P. Svenne,
 Int. J. Mod. Phys. E \textbf{19}, 1435 (2010).

 \bibitem{Hiyama00} E. Hiyama, M. Kamimura, T. Motoba, T. Yamada, and Y. Yamamoto,
 Phys. Rev. Lett. \textbf{85}, 270 (2000).

\bibitem{Yamada84} T. Yamada, K. Ikeda, H. Bando, and T. Motoba,
Prog. Theor. Phys. \textbf{71}, 985 (1984).

\bibitem{BIM83}H. Bando, K. Ikeda, and T. Motoba, Prog.
Theo. Phys. {\bf 69}, 918 (1983).

 \bibitem{lbl} J. Tauren and R. B. Firest, Evaluated Nuclear Structure Data File (ENSDF), http://ie.lbl.gov/TOI2003/index.asp.

\end{thebibliography}
\end{document}